\documentclass{aa}  

\usepackage{graphicx}
\usepackage{txfonts}
\usepackage{xcolor}

\begin{document}

   \title{Photometric redshift estimation of strongly lensed galaxies}


   \author{Danial Langeroodi\inst{1,2}
          \and
          Alessandro Sonnenfeld\inst{1}
          \and
          Henk Hoekstra\inst{1}
          \and
          Adriano Agnello\inst{2}
          }

   \institute{Leiden Observatory, Leiden University, Niels Bohrweg 2, 2333 CA Leiden, the Netherlands\\
   \email{danial.langeroodi@nbi.ku.dk}
   \and 
    DARK, Niels Bohr Institute, University of Copenhagen, Jagtvej 128, 2200 Copenhagen Ø, Denmark}

\date{\today}

 
  \abstract
   {
{Around $10^5$ strongly lensed galaxies are expected to be discovered with upcoming wide-field surveys such as Euclid and the LSST. Utilising these large samples to study the inner structure of lens galaxies requires source redshifts, which are needed to turn lens models into mass measurements. However, obtaining spectroscopic source redshifts for large samples of strong lenses is prohibitive with the current capacity of spectroscopic facilities. 
}
   
{
As an alternative to spectroscopy, we study the possibility of obtaining source photometric redshifts (photo-$z$s) for large samples of strong lenses. We pay particular attention to the problem of blending between the lens and the source light.
}
   
{
Our strategy consists of deblending the source and lens light by simultaneously modelling the lens galaxy and the background source in all available photometric bands, and then feeding the derived source colours to a template-fitting photo-$z$ algorithm.
We describe the lens and the source light with a S\'{e}rsic profile, and the lens mass with a singular isothermal ellipsoid.
We first test our approach on a simulated sample of lenses.
Then, we apply it to 23 real systems with broad-band photometry from the Hyper Suprime-Cam survey.
}

{ 
We identify the deviations of the lens light from a S\'{e}rsic profile and the contrast between the lens and source image as the main drivers of the source colour measurement error. Although the former is challenging to measure directly for real lenses, we find the latter to be sufficient for evaluating the accuracy of a measured source colour. We split the real sample based on the ratio $\Lambda$ of the lens to source surface brightness measured at the image locations. In the $\Lambda<1$ regime, the photo-$z$ outlier fraction is $20\%$, and the accuracy of photo-$z$ estimation is limited by the performance of the template-fitting process. In the opposite regime, the photo-$z$ outlier fraction is $75\%$, and the errors from the source colour measurements dominate the photo-$z$ uncertainty.
}

{
Measuring source photo-$z$s for lenses with $\Lambda<1$ poses no particular challenges compared to the isolated galaxy case. For systems with significant lens light contamination, however, improvements in the description of the surface brightness distribution of the lens are required: a single S\'{e}rsic model is not sufficiently accurate.
}
  }


   \maketitle
%

\section{Introduction}

Strong gravitational lensing is a rare phenomenon in which the close alignment of a background light source with a massive foreground lens results in multiple images of the former. 
The positions and surface brightness distribution of the multiple images tightly constrain the total projected mass (dark matter and baryonic component) of the lens.
In virtue of this feature, strong lensing has been used in the last two decades to investigate important aspects of the inner structure of massive galaxies, such as their total density profile \citep{T+K04,Aug++10,Bar++11,Bol++12,Son++13b}, their stellar mass-to-light ratio \citep{Tre++10,Son++15,SLC15,Son++19}, their dark matter density profile \citep{Son++12,New++15,O+A18,Sch++19,Sha++21}, and the presence of substructure \citep{Veg++10,Hez++16,Gil++20}.

One of the crucial pieces of information on which most strong lensing studies rely is the knowledge of the redshift of the lens and source galaxies; both are needed in order to convert angular measurements of image positions into measurements of mass.
Traditionally, these redshifts have been obtained by means of dedicated spectroscopic observations.
However, while lens galaxies are usually very bright, obtaining spectroscopic redshifts of strongly lensed sources can be relatively expensive, typically requiring approximately one hour of integration time on an eight-metre class telescope \citep{Ruf++11}. 
Moreover, a large fraction of lensed galaxies are found in the redshift range $1.5 < z < 3$, where they do not contribute with any strong emission line features to the optical part of the spectrum, thus requiring near-infrared observations \citep{Son++13b}.

The number of known strong lenses is growing rapidly, thanks to wide-field photometric surveys such as the Hyper Suprime-Cam Subaru Strategic Program \citep[HSC SSP,][]{Aih++18a}, the Kilo-Degree Survey \citep[KiDS,][]{Kui++15}, and the Dark Energy Survey \citep[DES,][]{DES16}, which are leading to the discovery of hundreds of lenses \citep{2018PASJ...70S..29S,Pet++19,Jac++19,2020A&A...642A.148S,Can++21,Roj++21,2022arXiv220602764S}.
However, the number of lenses with available spectroscopic data is only a fraction of the total. The largest sample of strong lenses with a spectroscopic measurement of the redshift of both the lens and the source galaxy is the Sloan ACS Lens Survey \citep[SLACS,][]{Bol++06}, which consists of $\sim100$ systems \citep{Aug++09}.
Although there are ongoing spectroscopic follow-up programmes \citep{2022AJ....164..148T}, this lack of spectroscopic observations is already limiting our ability to take advantage of the large number of currently known lenses, and will become an increasingly severe issue in the next decade; upcoming surveys such as Euclid\footnote{\url{https://www.euclid-ec.org/}}, the LSST\footnote{\url{https://www.lsst.org/}}, and the Chinese Space Station Telescope (CSST) are predicted to lead to the discovery of $10^5$ new strong lenses \citep{2015ApJ...811...20C}. While our capabilities for spectroscopic observations will also increase with upcoming facilities such as the 4-metre Multi-Object Spectrograph Telescope \citep[4MOST;][]{2019Msngr.175....3D}, the Maunakea Spectroscopic Explorer \citep{2019arXiv190404907T}, and MegaMapper \citep{2019BAAS...51g.229S}, most of these strong lenses will realistically not be followed up.

Photometric redshift (photo-$z$) measurements are a valid alternative to spectroscopic observations, when highly precise redshift estimates of individual galaxies are not strictly needed. This is the case for most strong lensing applications. The lensing deflection angle is a weak function of source redshift; therefore, redshift uncertainties do not propagate catastrophically to the mass measurements\footnote{Strictly speaking, this does not apply to lenses in which redshift of the lens is close to that of the source. However, such systems are rare because the lensing cross section decreases with decreasing source redshift.}.
Photometric redshift techniques have developed significantly in the last few years \citep[see][for a review]{SIH19}, especially because accurate photo-$z$s are a crucial ingredient for many experiments in cosmology \citep[see e.g.][]{Tan++18,Hil++21,Myl++21}.
Carrying out photometric redshift measurements on strong lenses, however, is generally more difficult than doing so on isolated objects.
The typical angular separation of a strongly lensed image from the  centre of its lens galaxy is $\sim1''$; as a result the lens and source light are often severely blended, especially when observed with ground-based imaging.
While this blending is usually a minor issue for the determination of the lens light distribution, it can have a strong impact on the ability to recover the magnitude and colours of the background source, which are needed to determine its photo-$z$.

In this work we investigate how accurately we can measure the  photo-$z$s of strongly lensed galaxies, when observed with photometric data similar to the HSC survey.
We chose the HSC survey because it is to date both the deepest and the one with the best image quality among the ground-based surveys used to find strong lenses. 
It also resembles the expected depth and image quality from LSST.
Our procedure for measuring the source photo-$z$ consists of a two-step process: first we fit the multi-band image of a strong lens with a simply parametrised model describing the lens light, the lens mass, and the source light; then we apply a template fitting-based photo-$z$ method to the measured magnitude and colours.
We first test this method on simulated lenses, then apply it to a set of 23 strong lenses from the Survey of Gravitationally-lensed Objects in HSC Imaging \citep[SuGOHI;][]{Son++19}.
We pay particular attention to our ability in measuring accurate colours.
For this purpose we investigate the importance of systematic errors associated with a non-perfect description of the surface brightness profile of the foreground galaxy.

Section \ref{data} covers the details of our sample of strongly lensed galaxies, consisting of both simulated and real systems. In Section \ref{methods} we explain our methodology for modelling the source and lens surface brightness and lens mass in multi-band images of galaxy-galaxy lens systems. In Section \ref{results: propagation of modelling uncertainties into colours} we identify the main drivers of the source colour measurement error by applying our method to the simulated sample. We estimate the source redshifts for the simulated and the real samples in Section \ref{results section}. We discuss and summarise our results in Sections \ref{discussion} and \ref{conclusion}.


\section{Data}
\label{data}

Our data consist of a real and a simulated sample of galaxy-galaxy strong lens systems, both based on imaging data in the broad-band $grizy$ photometry of the HSC survey. In this section we give a brief description of the HSC photometric data and describe our real and simulated samples. 

\subsection{HSC photometry}
\label{HSC photometry}

For our experiment we use $grizy$ imaging data from the second public data release of the HSC survey \citep[HSC PDR2;][]{2019PASJ...71..114A}.
For each object we obtain (from the HSC PDR2) a $101 \times 101$ pixel ($0.164\arcsec$ per pixel) sky-subtracted cutout (centred on the lens galaxy), a pixelised model of the point spread function (PSF), and the variance map, in each band. 
The median $i$-band seeing is $0.7\arcsec$ and its magnitude limit is $26.2$ ($5\sigma$ detection limit for a point source). 

The photometry of HSC is calibrated against the Pan-STARRS 1 (PS1) 3$\pi$ survey \citep{ps13pi}. 
At bright magnitudes, HSC coadd-measured star PSF fluxes have $\sim 0.03$ mag RMS residuals with respect to the PS1-based measurements and $\sim 0.04$ mag RMS residuals with respect to the SDSS-based \citep[SDSS DR9: the ninth data release of the Sloan Digital Sky Survey;][]{2012ApJS..203...21A} measurements \citep[see Figure 8 and Table 6 from][]{2018PASJ...70S...8A}. 
While the latter consists of both the absolute (the linear offset between the measured and true magnitudes) and relative (the standard deviation of the offset-corrected measured magnitudes around the true magnitudes) zero-point calibration inaccuracies, the former mostly consists of the relative calibration inaccuracies (since the zero-points of HSC are calibrated against the PS1).
Internal PS1 3$\pi$ comparisons and comparisons with the SDSS indicate that the photometry of both PS1 and SDSS surveys is accurate to $\sim 1\%$ in all bands \citep{2012ApJ...756..158S, 2016ApJ...822...66F}. 
However, because of spatial variations of the HSC filter transmissions, the RMS residuals of the HSC coadd-measured star PSF fluxes exceed the expected $\sim 1\%$ scatter inherited from the inaccuracies in PS1 and SDSS photometry \citep[see][]{2018PASJ...70...66K}.

Photo-$z$ estimation in this work (see Section \ref{methods: photometric redshift} for a detailed description of the method) is mostly sensitive to the colour measurement accuracy (as opposed to magnitude measurement accuracy), and therefore is only affected by the relative calibration accuracy. 
In general, a calibration of the colour-redshift relation compensates for this inaccuracy \citep[see e.g.][]{2010ApJ...709..644I}.
However, strong lenses are sufficiently rare that colours are measured from different pointings/observations, and thus are nonetheless affected by the relative calibrations. 
Therefore, the relative calibration accuracy ($\sim 0.03 \textnormal{ mag}$ for HSC) provides the lower threshold for the uncertainty on colour measurement of strongly lensed galaxies in HSC photometry.
The stage IV surveys (e.g. ESA-Euclid and Rubin-LSST) are expected to perform better than the HSC (we discuss this in more detail in Section \ref{real sample vs the simulated sample}).

\subsection{Real sample}
\label{real sample}

The real sample contains 23 galaxy-galaxy strong lenses from the study of \citet{Son++19}. All of these lenses consist of a foreground galaxy that belongs to the constant mass (CMASS) sample of the Baryon Oscillation Spectroscopic Survey \citep[BOSS;][]{2013AJ....145...10D}. They all have publicly available HSC $grizy$ imaging data, and spectroscopic measurements of both the source and the lens redshift. The lens redshifts range from 0.46 to 0.80, with a 0.58 median. The Einstein radii, measured by \citet{Son++19}, range from $0.72\arcsec$ to $2.14\arcsec$. The source redshifts cover a range between 0.94 and 3.12, with a median of 1.82 \citep{2018PASJ...70S..29S, 2018ApJ...867..107W, Son++19}. The average source-plane (i.e. de-lensed) half-light radii \citep[inferred from the best-fit models of][]{Son++19} range from $0.05\arcsec$ to $0.95\arcsec$, with a $0.20\arcsec$ median. The source $i$-band magnitudes are between 22.86 and 27.02.

\subsection{Simulated sample}
\label{simulations}

We simulated each lens system in our mock sample by taking the five-band $grizy$ HSC images of a real massive galaxy as the lens and adding the simulated images of a lensed source, closely following the method that \cite{2020A&A...642A.148S} used to produce the training sample for the citizen science lens finding programme  Space~Warps-HSC\footnote{\url{https://www.spacewarps.org}}. We selected galaxies from the CMASS sample for this purpose since this is the sample from which the foreground galaxies of our real lenses are drawn (see Section \ref{real sample}). In particular, we drew elliptical galaxies from the subset of visually selected  CMASS galaxies with stellar mass above $10^{11} M_{\odot}$ of \citet{2019A&A...622A..30S}.

Modelling the surface brightness distribution of the simulated lensed source in the lens plane requires associating a lensing potential with the foreground galaxy. We assign a singular isothermal ellipsoid \citep[SIE,][]{1994A&A...284..285K} mass profile to each lens, constructed from aligned co-centric elliptical isodensity contours of equal axis ratio ($q_{\textnormal{SIE}}$). \cite{2008ApJ...682..964B} showed that the SIE mass model can provide accurate reproductions of the source light distribution in the lens plane. The SIE is a five-parameter model fully described by the position of its centroid pixel ($x_{\textnormal{centroid}}$ and $y_{\textnormal{centroid}}$), axis ratio ($q_{\textnormal{SIE}}$), position angle ($p_{\textnormal{A}}$), and angular Einstein radius ($b_{\textnormal{SIE}}$; defined as the circularised radius of the isodensity contour that encloses an average surface mass density equal to the critical surface mass density for lensing). 

We drew the angular Einstein radii from a truncated normal distribution between 0.5'' and 3.0'', centred on 1.5'', and with a 0.5'' width. We set the lower limit because systems with angular Einstein radii significantly smaller than the seeing of HSC are difficult to detect as strong lenses and are usually missed by photometry-based lens searches. The upper limit is imposed because in this work we are mostly interested in galaxy-scale lenses as opposed to group-scale lenses, which dominate the image separations larger than $3\arcsec$ \citep{2006MNRAS.367.1241O}.

We do not intend to align the lens mass profile to its surface brightness distribution because our aim is to span a broad range of parameter space with our simulation. Therefore, we randomly draw the axis ratios between 0.4 and 1.0 and position angles between $0^{\circ}$ and $180^{\circ}$, regardless of the lens light. As we explain in Section \ref{results: propagation of modelling uncertainties into colours}, if we can reproduce the strong lensing image configuration in this general setting (where the lens mass and light profiles are not aligned), we should also be able to reproduce them in the more realistic case where the lens mass and light profiles are closely aligned. Furthermore, we select the polar coordinates ($r, \phi$) of the SIE centroids randomly in one-pixel radius circles centred on the centre of the lens cutout in each photometry band (i.e. $x= 50$, $y = 50$ pixels). In other words, we randomly draw $r_{\textnormal{centroid}}^2$ between 0 and $1''^2$ and the $\phi_{\textnormal{centroid}}$ between 0 and $2\pi$. 

\begin{figure}
    \centering
    \includegraphics[width=9cm]{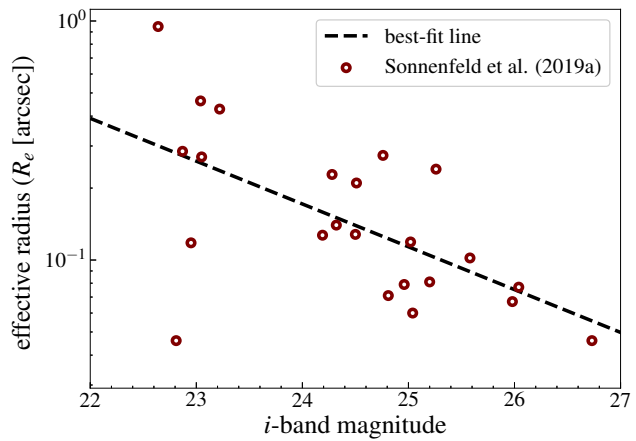}
    \caption{Effective radius ($R_{\mathrm{e}}$ [arcsec]) plotted against the $i$-band magnitude ($m_i$) for the sources in our real sample of strongly lensed galaxies. The dashed line shows the best-fit line, given by Eq.~(\ref{effective radius}). This magnitude-size relation (with scatter) is used to assign the effective radii corresponding to the drawn $i$-band magnitudes of the source galaxies in our simulated sample of strongly lensed galaxies.}
    \label{effective_radius_plot}
\end{figure}

We describe the surface brightness of source galaxies in the source plane with elliptical S\'{e}rsic surface brightness profiles, constructed from co-centric elliptical isophotes \citep{1968adga.book.....S}. The S\'{e}rsic light profile is defined as  
\begin{equation}
    I(x,y) = I_0\exp\bigg\{-b(n)\left(\frac{R}{R_{\mathrm{e}}}\right)^{1/n}\bigg\},
    \label{eq:remag}
\end{equation}
where $I_0$ is the amplitude, $n$ is the S\'{e}rsic index describing the central concentration of the light profile, and 
\begin{equation}
    b(n) \approx 2n-\frac{1}{3}+\frac{4}{405n}+\mathcal{O}(n^{-2})
\end{equation}
is defined such that the effective radius ($R_{\mathrm{e}}$) encloses half the light \citep{1999A&A...352..447C}.

\begin{figure*}
    \centering
    \includegraphics[width=\hsize]{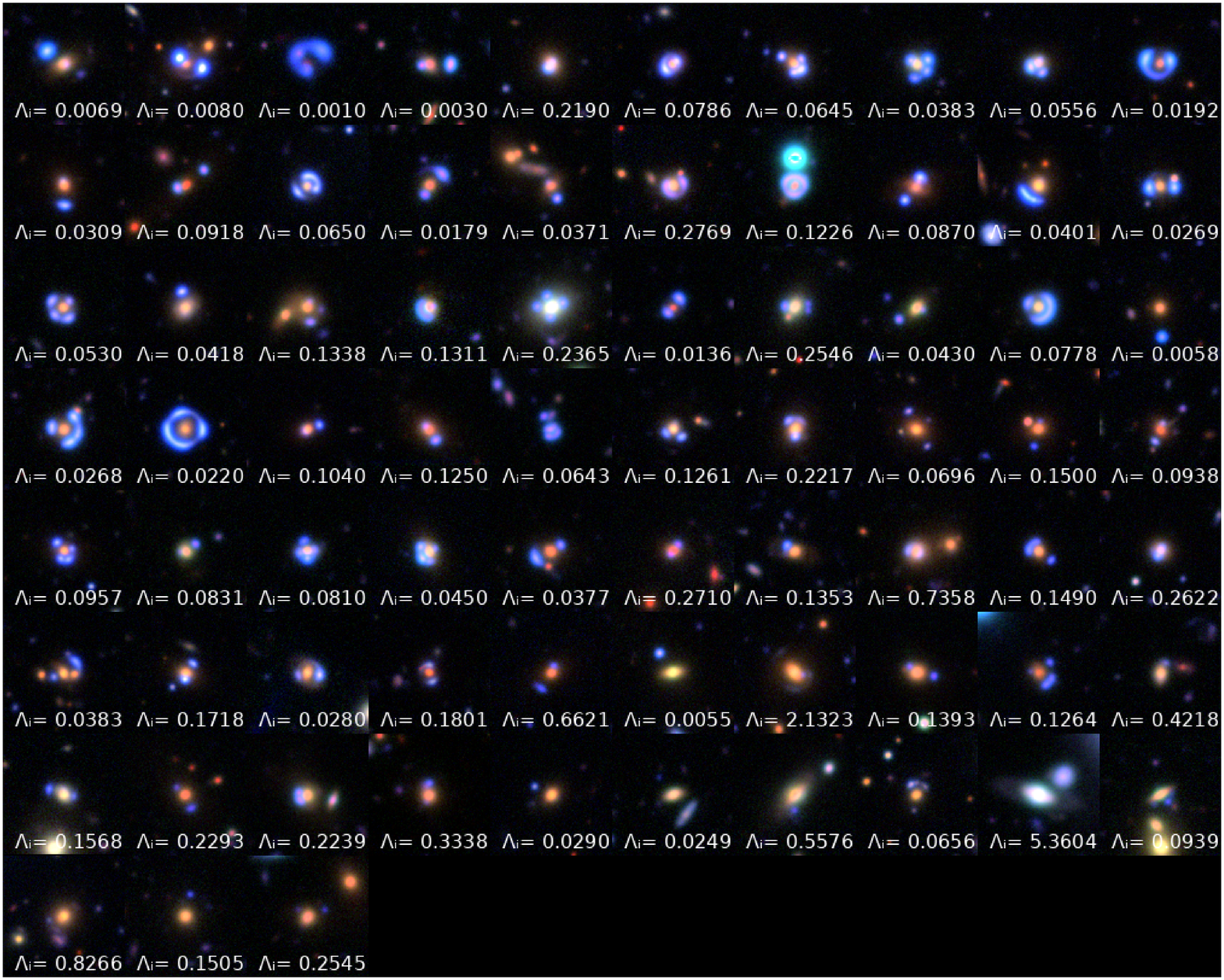}
    \caption{$101 \times 101$ pixel ($0.164\arcsec$ per pixel) $gri$ colour-composite images of the 73 systems in our simulated sample of strongly lensed galaxies. Source $i$-band magnitudes increase from 20.50 to 25.90 (from left to right and top to bottom). 
    For each system, the $i$-band ratio $\Lambda_i$ of the lens to source surface brightness measured at the image locations (introduced and discussed in Section \ref{section: the accuracy of the lens light model}) is overwritten on its image.}
    \label{simulated sample}
\end{figure*}

We assume that the colours of the background source are spatially uniform.
Therefore, for each simulated source, we use the same values of structural parameters (axis ratio, position angle, S\'{e}rsic index, and half-light radius) in all photometric bands. We randomly draw the axis ratios of the co-centric elliptical isophotes between 0.3 and 1.0, and the position angles between $0^{\circ}$ and $180^{\circ}$. 
We draw the S\'{e}rsic indices randomly between 0.5 and 2 because the detected lensed sources are typically late-type galaxies, best fit by a S\'{e}rsic index between 0.5 and 2 \citep[see e.g.][]{2009ApJS..182..216K, 2012MNRAS.421.1007K}.
The prevalence of late-type galaxies among the lensed source population is partly intrinsic and partly a result of selection effects: blue late-type galaxies can be more easily detected in photometry-based search algorithms as they can be more easily separated from the light of the lens galaxy, which is typically red \citep[see e.g.][]{Gav++14,2018PASJ...70S..29S}.

We assigned effective radii assuming a magnitude--size relation, with scatter, which we inferred from the \citet{Son++19} sample (Figure \ref{effective_radius_plot}).
In particular, we fitted the unlensed $i$-band magnitude $m_i$ and source effective radius $R_{\mathrm{e}}$ with the  model 
%
\begin{equation}
    \log_{10}(R_{\mathrm{e}}) \sim -0.179 m_i + 3.533 + \mathcal{D}(0,0.272),
    \label{effective radius}
\end{equation}
where $\mathcal{D}(0,0.272)$ is a normal distribution centred at 0 with a dispersion equal to the standard deviation of the best-fit line (0.272). Motivated by the distribution of source magnitudes in the sample of real strongly lensed galaxies \citep[Section \ref{real sample} and Table 4 in][]{Son++19}, we draw the source $i$-band magnitudes from a normal distribution centred on 24.4~mag, with a 1.2~mag dispersion. We assign effective radii corresponding to the drawn $i$-band magnitudes using Equation \ref{effective radius}.

The first requirement for producing strongly lensed images is to place the source within the region that is mapped into multiple images, which for an SIE mass model corresponds to the radial caustic and has a circularised radius similar to the value of the Einstein radius. However, placing the source too close to the radial caustic produces three images,  two of which are very faint and close to the deflector, and one that has minimal distortion. In practice, these image configurations cannot be identified as strongly lensed. To prevent the production of such image configurations, following \cite{2020A&A...642A.148S}, we assign $r_{\textnormal{source}}$ by drawing $x$ from the $p(x) \propto x\exp(-4x)$ distribution, where $x = r_{\textnormal{source}}/b_{\textnormal{SIE}}$. Multiplying the drawn $x$ by the previously drawn Einstein radius ($b_{\textnormal{SIE}}$) gives the $r_{\textnormal{source}}$. We draw the $\phi_{\textnormal{source}}$ randomly between 0 and $2\pi$. For each drawn source position $(r_{\textnormal{source}},\phi_{\textnormal{source}})$, we calculate the number and magnifications of the strongly lensed images. 
If the drawn source position results in multiple images, we accept it. Otherwise, we draw a new source position and repeat the same procedure. 

Unlike its structural parameters and position, the amplitude of the source S\'{e}rsic profile varies with photometry band. This reflects the spectral energy distribution (SED) of galaxy, which depends on both its redshift and spectral type. We draw the source redshifts uniformly between 1.0 and 3.0, while selecting the spectral types randomly from the Bayesian photometric redshift \citep[BPZ;][]{2000ApJ...536..571B} SED templates \citep{1980ApJS...43..393C,1996ApJ...467...38K,1997AJ....113....1S,1999ApJ...513...34F,2006AJ....132..926C}. 
Since the majority of source galaxies in detected strongly lensed samples are late-type galaxies, we only draw late-type spectral types from the BPZ SED templates.
Each drawn redshift--spectral type pair corresponds to an SED continuum, which we convert to multi-band $grizy$ magnitudes, using the $grizy$ response functions. We re-scale the calculated $grizy$ magnitudes, such that the $i$-band magnitude matches its previously drawn value ($m_i$).

Provided a source position, source light profile, and a lens mass profile, we calculate the surface brightness distribution of the strongly lensed image in each photometric band. For each band, we convolve the resulting image with the PSF of the real lens galaxy, and add the PSF-convolved lensed image to the lens image.

We sift through our simulated sample to discard the cases that would not be detected as strongly lensed galaxies with conventional detection techniques. This is the case when the lensed source is not sufficiently bright compared to the background sky noise, $\sigma_{\text{sky}}$. We define the source footprint as the pixels in the source-only image with pixel values larger than $3\sigma_{\text{sky}}$. Since the sources in \cite{Son++19} are typically brightest in $g$-band, this band often has the largest number of source region pixels. We assume that detection in one band ($g$ here) is sufficient to identify the system as strongly lensed. Therefore, we discard any simulated source for which the number of $g$-band source region pixels is smaller than 20 as undetectable.  We make a sample of 73 galaxy-galaxy strong lens systems in the $grizy$ HSC broad-band photometry. Figure \ref{simulated sample} shows the HSC $gri$ colour-composite images of these systems. 


\section{Methods}
\label{methods}

In this section we describe our method for measuring the source photometry and using the measured photometry to estimate the source photometric redshift (photo-$z$). Source photometry is measured by deblending the lens and source light, which is done by modelling the lensed systems (Section \ref{modelling the strongly lensed systems}). We use a photometric redshift estimation algorithm (photo-$z$ algorithm) to convert the measured colours to redshift probability distribution functions (PDFs). We describe the photo-$z$ algorithm used in this work in Section \ref{methods: photometric redshift}.

\subsection{Modelling the lensed system}
\label{modelling the strongly lensed systems}

Our lens model was designed to reproduce the image configuration of the strongly lensed system by simultaneously modelling the lensing potential, the surface brightness distribution of the foreground lens galaxy, and the  unlensed surface brightness distribution of the source galaxy. 
We describe them with a SIE mass profile, an elliptical S\'{e}rsic profile in the image plane, and an elliptical S\'{e}rsic profile in the source plane, respectively.

For both the lens and the source galaxy, we assume that the  structural parameters of the light profile (axis ratio, S\'{e}rsic index, half-light radius, position angle, and centroid position) are the same in each band.
Additionally, we allow  the four colours to vary independently of each other. For the five-band HSC $grizy$ photometry used in this work, the colours  are $g-i$, $r-i$, $i-z$, and $i-y$. 
For each set of source and lens light profile parameters ($2 \times 9$ parameters), and lens mass profile parameters (five parameters), we find the source and lens $i$-band magnitudes that minimise
\begin{equation}
    \chi^2 = \sum_{\lambda} \sum_{j} \bigg( \frac{I_{\lambda,j}^\text{mod, L} + I_{\lambda,j}^\text{mod, S} - I_{\lambda,j}^\text{obs}}{\sigma_{\lambda,j}^2} \bigg)\;,
    \label{chi squared}
\end{equation}
in which $I_{\lambda,j}^\text{mod, L}$ ($I_{\lambda,j}^\text{mod, S}$) is the $\lambda$-band lens (source) PSF-convolved S\'{e}rsic model, evaluated at the $j$'th pixel, $I_{\lambda,j}^\text{obs}$ is the observed value of the $j$'th pixel, and $\sigma_{\lambda,j}$ is the observational uncertainty provided by the variance file. 

We use the EMCEE package \citep{2013PASP..125..306F} to explore the parameter space with a Markov chain Monte Carlo (MCMC) sampler, searching for the parameter values that minimise the $\chi^2$ in Equation \ref{chi squared}. 
Since we are only interested in obtaining a good description of the system in the area that includes the multiple images of the source, we only fitted data within a circular region with a 20 pixel radius ($3.36\arcsec$) around the lens centroid in each photometric band. 

In addition to the lens and the source, image cutouts typically contain contaminating bright objects. These are either satellite galaxies and extended features in the lens plane or accidental alignments of other bright objects that do not belong to the lens or source planes. 
These contaminants can bias the measurements if not taken into account.
If the contaminant is sufficiently faint and not closely aligned with the lensed system, its surface brightness at the lens or source image position is negligible. Hence, it is sufficient to discard the pixels at which the contaminant dominates the image cutout. To do this we identify the contaminants by running the Source Extractor \citep[{\tt SExtractor};][]{1996A&AS..117..393B} on the $i$-band cutouts. 
We exclude (i.e. mask) these segments from the chi-squared optimisation by excluding their pixels from the sum in Equation \ref{chi squared}. 

\begin{figure}
    \centering
    \includegraphics[width=9cm]{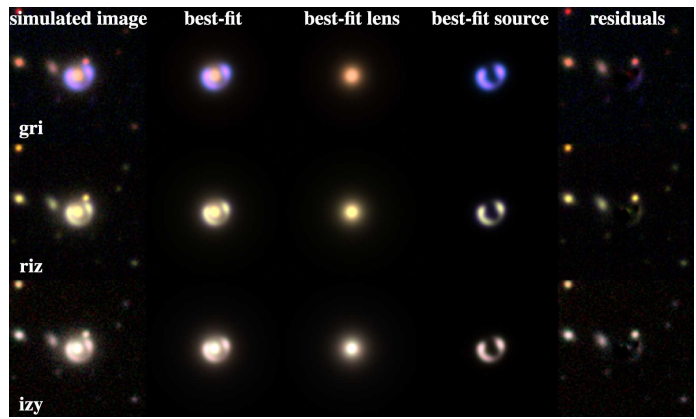}
    \caption{Example of a two-component best-fit model to the $grizy$ images of a simulated strongly lensed galaxy-galaxy system. From top to bottom: Colour-composite $gri$, $riz$, and $izy$ images. The first column from the left shows the simulated image. The second column shows the PSF-convolved best-fit image. The third and fourth columns show the best-fit lens-only and the best-fit source-only light profiles. The last column shows the residuals (best-fit-subtracted original image). The contaminants were masked by running the {\tt SExtractor} on the $i$-band image.}
    \label{fitted multiband: simulations}
\end{figure}

However, if the contaminant is too bright or closely aligned with the lensed system, its surface brightness at the lens or source image position becomes significant. This additional contaminating light results in systematic errors in modelling the lensed system. Moreover, often {\tt SExtractor} cannot separate these contaminants from the lensing system. We flagged these cases by running {\tt SExtractor} on the original lens image (for the simulated sample) and by visual inspection (for both the simulated and the real samples). In these cases we modelled the contaminant light explicitly, with an additional S\'{e}rsic component.
Figure \ref{fitted multiband: simulations} shows an example of a fitted model to the $grizy$ images of a simulated strong lens system.

\subsection{Source photometric redshift estimation}
\label{methods: photometric redshift}

We use BPZ \citep{2000ApJ...536..571B} to obtain photo-$z$ estimates from the measured magnitudes. 
BPZ fits the PSF-corrected multi-band magnitude measurements of the galaxy surface brightness with a grid of SED templates varying in both redshift $z$ and spectral type $T$. The outcome is a probability distribution $P(z,T)$. 
The marginal posterior probability in redshift, $P(z)$, is calculated by marginalising over the spectral type $T$.
BPZ contains elliptical, spiral, irregular, and starburst spectral types, including the templates from \cite{2000ApJ...536..571B} \citep{1980ApJS...43..393C,1996ApJ...467...38K,1997AJ....113....1S,1999ApJ...513...34F}, re-calibrated in \cite{2004ApJS..150....1B} with the addition of two bluer templates of \cite{2006AJ....132..926C}, all corrected for intergalactic absorption according to \cite{1995ApJ...441...18M}. These are the same spectral types that we used for the simulations (Section \ref{simulations}).

\begin{figure*}
    \centering
    \includegraphics[width = \hsize]{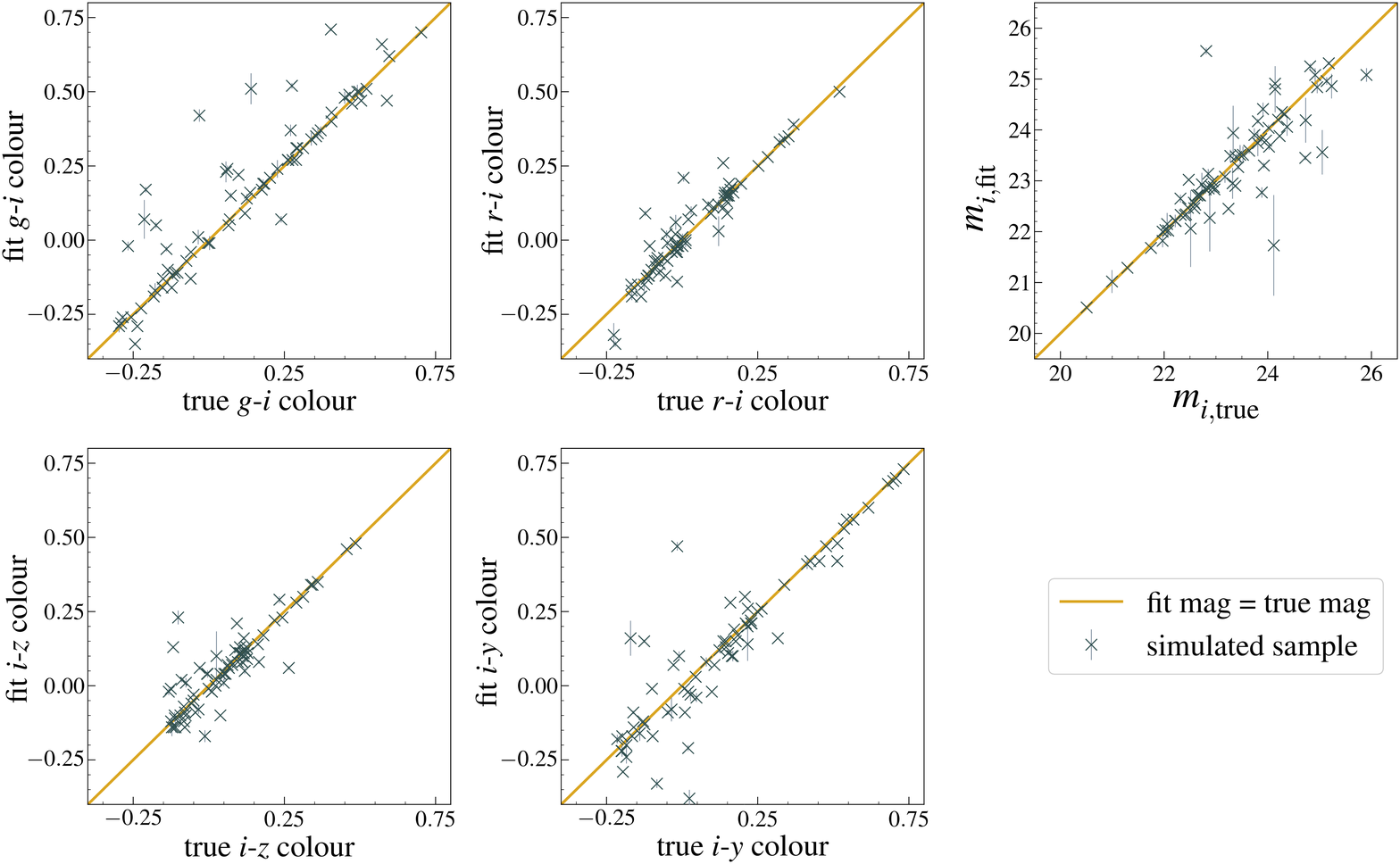}
    \caption{Best-fit vs true $g-i$, $r-i$, $i-z$, and $i-y$ source colours (four left panels), and the best-fit vs true source $i$-band magnitudes (top right panel), for 73 strongly lensed galaxy-galaxy systems in our simulated sample. The dark yellow lines indicate where the fitted values equal the true values. The error bars show the $1\sigma$ uncertainties inferred from the MCMC chains. The best-fit $i$-band magnitudes have a 0.07 bias around their true values and a 0.29 standard deviation. The best-fit $g-i$, $r-i$, $i-z$, and $i-y$ colours have 0.007, $-0.002$, $-0.006$, and $-0.015$ bias around their true values and 0.03, 0.02, 0.03, and 0.06 standard deviations, respectively. The colour measurement is much more accurate than the magnitude measurement.}
    \label{magnitudes and colors for fits: simulations}
\end{figure*}

The BPZ adopts a magnitude-dependent prior that gives the redshift and spectral type probability distribution for a reference-band magnitude. \cite{2000ApJ...536..571B} calibrated this prior empirically on the spectroscopic redshifts and spectral type classifications of the Hubble Deep Field North observations \citep[HDF-N;][]{1996AJ....112.1335W} and Canada-France Redshift Survey \citep[CFRS;][]{1995ApJ...455...50L}.
For simulated systems, instead of this prior we adopt a flat prior that is truncated below the lens redshift and above a maximum value of $z=6$. 
This is appropriate because the source redshifts in our simulated sample are not correlated with their reference-band magnitudes; this choice was made to span a broad range of parameter space (similar to the choice made in Section \ref{simulations} for not aligning the lens mass profile to the lens light).
In the case of our 23 real strongly lensed galaxies, using either the same prior as  used for the simulated sample or the magnitude-dependent prior of BPZ, truncated below the lens redshift, does not change the accuracy of photo-$z$ estimation.

\section{Systematics in source colour measurement}
\label{results: propagation of modelling uncertainties into colours}

In this section we identify the main drivers of the source colour measurement error. 
We do this by assessing the accuracy of the method of Section \ref{modelling the strongly lensed systems} in measuring the source colours of our simulated sample. 
We measure the source colours and evaluate the uncertainties and biases of these measurements (Section \ref{colour measurement}). We hypothesise that the colour measurement errors are mainly correlated with the deviations of the lens light from a S\'{e}rsic profile and with the contrast between the lens and source image. We investigate the former in Section \ref{section: the accuracy of the lens light model} and the latter in Section \ref{sect:contrast}.

\subsection{Colour measurement}
\label{colour measurement}

Following the method of Section \ref{modelling the strongly lensed systems}, we measure the source $i$-band magnitude and colours for each system in our simulated sample. The right-hand panel in Figure \ref{magnitudes and colors for fits: simulations} shows the best-fit source $i$-band magnitudes plotted against the true values. The error bars show the $1\sigma$ uncertainties inferred from the MCMC chains. The best-fit values have a $-0.011$ bias with respect to the true values and a 0.29 standard deviation. The inaccuracy of the $i$-band magnitude measurements is largely due to the lens model degeneracies: there is not a unique lens mass model that can reproduce a given strong lensing image configuration. In particular, the best-fit source magnitudes are degenerate with the best-fit lens mass model, resulting in large magnitude measurement inaccuracies. 

The remaining four panels in Figure \ref{magnitudes and colors for fits: simulations} show the $g-i$, $r-i$, $i-z$, and $i-y$ best-fit source colours plotted against their true values. The error bars show the statistical $1\sigma$ uncertainties inferred from the MCMC chains. The best-fit $g-i$, $r-i$, $i-z$, and $i-y$ source colours have 0.007, $-0.002$, $-0.006$, and $-0.015$ bias around their true values and 0.03, 0.02, 0.03, and 0.06 standard deviations, respectively. The colour measurement is much more accurate than the magnitude measurement  because gravitational lensing is achromatic, which is why lens model degeneracies do not affect the colour measurement. The colour measurement errors are, on average, as low as the zero-point systematic uncertainty of the HSC \citep[see e.g. Section \ref{HSC photometry}, and][]{2019PASJ...71..114A}.


The formal uncertainties inferred from the MCMC chains significantly underestimate the errors on the magnitude and colour measurements. 
In other words, the $1\sigma$ error bars of each plot in Figure \ref{magnitudes and colors for fits: simulations} (0.12 for the $i$-band magnitude and 0.011, 0.009, 0.007, and 0.008, respectively for the $g-i$, $r-i$, $i-z$, and $i-y$ colours) are significantly smaller than the standard deviation of the corresponding best-fit measurements around their true values (as measured earlier: 0.29 for the $i$-band magnitude and 0.03, 0.02, 0.03, and 0.06, respectively, for the $g-i$, $r-i$, $i-z$, and $i-y$ colours). 
This suggests that non-negligible systematic uncertainties are present in magnitude and colour measurement. 
Although the colour measurement errors and biases are significantly smaller than those of the magnitude measurement ($i$-band here), they can translate into significant systematics in photo-$z$ estimation. 
In the following section, we investigate the main drivers of source colour measurement inaccuracy.

\subsection{Accuracy of the lens light model}
\label{section: the accuracy of the lens light model}


The cause for the biases in the measured colours must lie in the differences between the true surface brightness distribution of the lens systems and that of the model fitted to them. We  describe the source galaxies with the same model family used to generate them, a S\'{e}rsic profile, and therefore that aspect of the model is accurate by construction. In contrast, the lens galaxies in our simulated sample are real objects whose surface brightness in general deviates from a S\'{e}rsic model. 
These deviations cause positive or negative lens light residuals in the model-subtracted images. The source light model overfits the leftover light at the source image position to minimise the $\chi^2$ (Equation \ref{chi squared}). Since the lens colours (and hence, the colours of their residuals) differ from the source colours, this overfitting results in  best-fit source colours that deviate from the true values. 

In this section we investigate how deviations of the lens light from a S\'{e}rsic profile affect the source colour measurement. 
First, we derive the expected correlation between the lens residuals at the source image position and the source colour measurement error by assuming that all the lens residuals at the source image position get overfitted by the best-fit source light model. We then use this correlation to derive the expected colour measurement error in terms of the deviations of the lens light from a S\'{e}rsic profile and the contrast between the lens and source image. 
We focus on the $g-i$ colour, $C_{gi}$, but the equations for the  other colours are identical.

Since lensing is achromatic, in the absence of lens residuals the source $g-i$ colour is simply the difference between the $g$-band and $i$-band magnitudes of the source image\footnote{Throughout this paper the `source image' refers to the image at the lens plane.}
\begin{equation}
    C_{gi} = -2.5(\log_{10} I_0^g - \log_{10} I_0^i)\;,
\end{equation}
where $I_0^{\textnormal{band}}$ is the source image flux in each band. In the presence of lens residuals, however, the measured colour becomes 
\begin{equation}
    C_{gi}^{\prime} = -2.5\bigg[ \log_{10} (I_0^g+\delta I^g) - \log_{10} (I_0^i+\delta I^i) \bigg]\;,
\end{equation}
where $\delta I^{\textnormal{band}}$ is the lens residual flux at the source image position. It is straightforward to show that the source colour measurement error ($\delta C_{gi} = C_{gi}^{\prime}-C_{gi}$) is given by
\begin{equation}
    \delta C_{gi} = -2.5 \log_{10}\frac{1+\psi_g}{1+\psi_i}\;,
    \label{equation: omega vs delta c}
\end{equation}
where we define
\begin{equation}
    \psi_{\textnormal{band}} = \frac{\delta I^{\textnormal{band}}}{I^{\textnormal{band}}_0}. 
    \label{psi text}
\end{equation}
In Appendix \ref{sect:lensres} we measure the same correlation from the best-fit models of Section \ref{colour measurement} and compare it with the expected correlation from Equation \ref{equation: omega vs delta c}.

\begin{figure}
    \centering
    \includegraphics[width = 9cm]{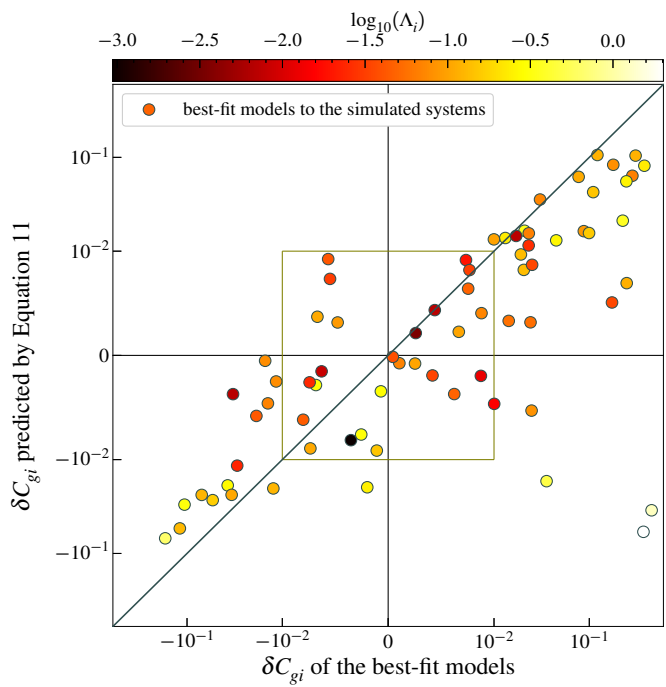}
    \caption{Source $g-i$ colour measurement error $\delta C_{gi}$ as  predicted by inserting the $\alpha_i$, $\alpha_g$, $\Lambda_i$, and $\Lambda_g$ ($i$-band and $g$-band deviation of the lens light from a S\'{e}rsic profile, and  $i$-band and $g$-band ratio of the lens to source surface brightness measured at the image locations, respectively) of the best-fit models into Equation \ref{eq:req} (derived correlation between $\delta C_{gi}$, $\alpha$, and $\Lambda$), plotted against the true $\delta C_{gi}$ of the best-fit models.
    The diagonal solid black line indicates were the predicted and true $\delta C_{gi}$ are equal. The data point for each simulated system is colour-coded with its $\Lambda_i$ value. The data points are plotted linearly within the inner olive box.}
    \label{deltaC vs deltaC plot}
\end{figure}

We then define the deviation of the lens light from a S\'{e}rsic profile at the source image position, $\alpha$, as
\begin{equation}
    \alpha_{\textnormal{band}} = \frac{ I_{\textnormal{res}}^{\textnormal{band}} (x_{c},y_{c}) }{ I_{\textnormal{S\'{e}rsic}}^{\textnormal{band}} (x_{c},y_{c}) }\;.
\end{equation}
In each band, $I_{\textnormal{res}}^{\textnormal{band}} (x_{c},y_{c})$ is the lens residual flux and $I_{\textnormal{S\'{e}rsic}}^{\textnormal{band}} (x_{c},y_{c})$ is the best-fit lens light model flux, both at a $3\times3$ pixel cutout $(x_{c},y_{c})$ centred on the pixel that is brightest in the $i$-band (lensed) source image.
Moreover, we introduce the $\lambda$-band contrast between the lens and source, $\Lambda_{\lambda}$, defined as the $\lambda$-band ratio of the best-fit lens light model to the best-fit (lensed) source light model at the $(x_{c},y_{c})$ pixels ($\Lambda_i$ of each system in our simulated sample is overwritten on its $gri$ colour-composite image in Figure \ref{simulated sample}). With the above definitions, $\psi_{\textnormal{band}}$ in Equation \ref{psi text} can be re-written as 
\begin{equation}
    \psi_{\textnormal{band}} = \alpha_{\textnormal{band}}\Lambda_{\textnormal{band}}\;.
    \label{psi new}
\end{equation}
Using Equation \ref{psi new} to insert for $\psi_g$ and $\psi_i$, Equation \ref{equation: omega vs delta c} becomes
\begin{equation}
    \delta C_{gi} = -2.5 \log_{10} \frac{1+ \alpha_g \Lambda_g}{1+\alpha_i \Lambda_i}\;.
    \label{eq:req}
\end{equation}

\begin{figure}
    \centering
    \includegraphics[width = 9cm]{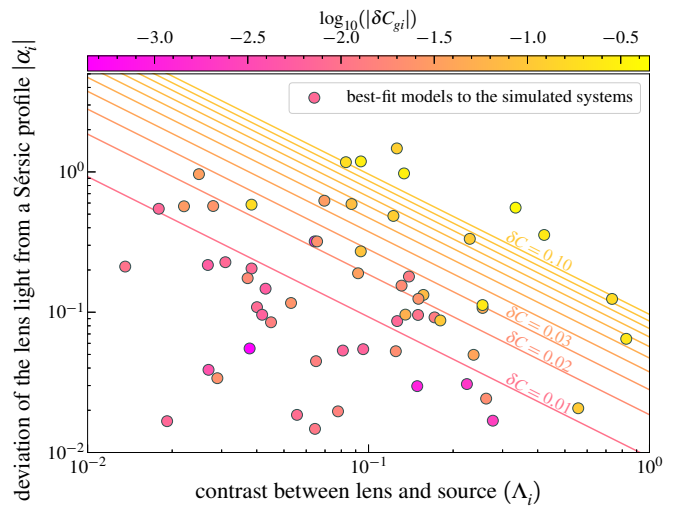}
    \caption{Highest allowed deviation of the lens light from a S\'{e}rsic profile ($\alpha_i$) as a function of the $i$-band contrast between the lens and source image ($\Lambda_i$) to achieve source colour measurements that are more accurate than a set threshold. The solid lines are generated from the derived correlation between $\delta C_{gi}$, $\Lambda_i$, and $\alpha_i$ (Equation \ref{eq:alphasimp}). Each solid line shows $\alpha_i$ as a function of $\Lambda_i$, for a constant $\delta C_{gi}$. The value of $\delta C_{gi}$ increases from 0.01 to 0.1, in 0.01 intervals (from bottom to top). 
    The scattered data points show the distribution of $\alpha_i$ and $\Lambda_i$ for the best-fit models to the simulated systems. The scattered data points are colour-coded with the $\delta C_{gi}$ values of the best-fit models. In agreement with the derived correlation (Equation \ref{eq:alphasimp}), the $\delta C_{gi}$ of the best-fit models increases perpendicularly to the solid constant $\delta C_{gi}$ lines. This figure indicates that $\Lambda_i$ is the main parameter that determines the maximum allowed $\alpha_i$ to have $\delta C_{gi}$ below a desired threshold.}
    \label{alpha vs lambda plot}
\end{figure}

In order to verify the validity of Equation \ref{eq:req}, we use it to calculate the expected $\delta C_{gi}$ for each set of $\alpha_i$, $\alpha_g$, $\Lambda_i$, and $\Lambda_g$ measured for the simulated systems, and compare it with the true $\delta C_{gi}$ of the best-fit models (Section \ref{colour measurement}). 
Figure \ref{deltaC vs deltaC plot} shows the close agreements between the predicted $\delta C_{gi}$ and the true $\delta C_{gi}$ of the best-fit models. 
The $1\sigma$ scatter of the $|\delta C_{gi}|$ predicted with Equation \ref{eq:req} is 0.02 and their bias is 0.002, both in close agreement with the $g-i$ colour measurement error and bias found in Section \ref{colour measurement} (0.03 and 0.007, respectively)

\begin{figure}
    \centering
    \includegraphics[width=9cm]{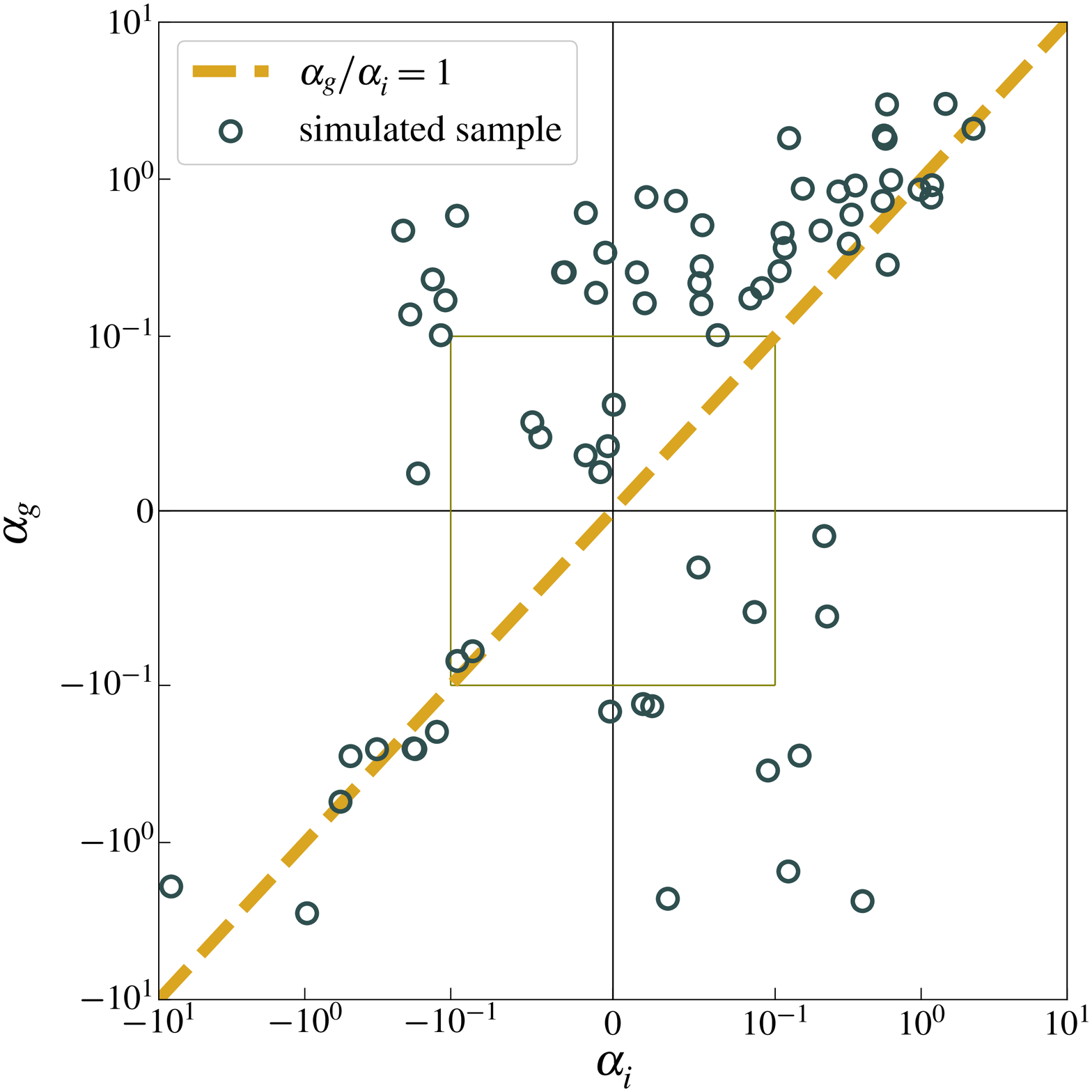}
    \caption{Lens light residuals can have non-negligible colour gradients. This figure shows the distribution of the $g$-band and the $i$-band deviations of the lens light from a S\'{e}rsic profile ($\alpha_g$ and $\alpha_i$) for the systems in our simulated sample. The dashed line indicates where residuals have no colour gradients (i.e. where $\alpha_g$ = $\alpha_i$). The data is plotted linearly within the inner olive box. This figure shows that the deviations of the lens light from a S\'{e}rsic profile in different bands are not necessarily correlated.}
    \label{agai}
\end{figure}

Equation \ref{eq:req} can also be used to calculate the required accuracy of the lens light model (i.e. the maximum allowed $\alpha$) as a function of $\Lambda$, to achieve source colour measurements that are more accurate than a desired threshold.
For HSC-like data, a reasonable choice for the colour error threshold is $\delta C_{gi} \approx 0.03$ since this is the typical zero-point systematic uncertainty of HSC photometric measurements.
For simplicity, we do this in the limiting case where $\Lambda_g$ approaches 0, which allows us to reduce the dimensionality of the problem. This is a good approximation for our lens set since the lens galaxy is typically very faint in the $g$-band, compared to the source (see Section \ref{methods: photometric redshift}, Figure \ref{simulated sample}, and Figure 1 in \citealt{Son++19}).
We write $\alpha_i$ in terms of $\delta C_{gi}$ and $\Lambda_i$:
\begin{equation}
    \alpha_i = \frac{10^{\delta C_{gi}/2.5}-1}{\Lambda_i}\;.
    \label{eq:alphasimp}
\end{equation}
This formulation is particularly useful for understanding the correlations between $\delta C_{gi}$ and $\alpha_i$, and $\delta C_{gi}$ and $\Lambda_i$, independently. The latter is of particular importance because $\Lambda_i$ can be measured on real strongly lensed systems, while $\alpha_i$ is challenging to measure without the lens-only image (which is only available for the simulated sample).

Equation \ref{eq:alphasimp} shows that large deviations from a S\'{e}rsic profile (large values of $\alpha_i$) can potentially lead to large errors on the colour. The impact of $\alpha_i$, however, is modulated by the lens-source contrast $\Lambda_i$. For large values of $\Lambda_i$, for example, even small values of $\alpha_i$ can propagate to large values of $\delta C_{gi}$, and vice versa, for $\Lambda_i \approx 0$ the impact of the lens light contamination becomes negligible.

Figure \ref{alpha vs lambda plot} shows the required $\alpha_i$ as a function of $\Lambda_i$ for a set of values of $\delta C_{gi}$ threshold (solid lines).
%
The error $\delta C_{gi}$ increases from 0.01 to 0.1 from the bottom to the top in 0.01 intervals. The scattered data points show $\alpha_i$ versus $\Lambda_i$ for the best-fit models to the simulated systems, colour-coded with their $\delta C_{gi}$. 
As predicted, $\delta C_{gi}$ of the best-fit models increase perpendicularly to the $\delta C_{gi}$ = constant lines (i.e. the same direction in which the predicted $\delta C_{gi}$ of the solid lines increase). 
Figure \ref{alpha vs lambda plot} indicates that for a realistic range of lens light parameters and colours, such as those spanned by our simulation, the contrast between the lens and source $\Lambda_i$ is the main parameter that determines the required accuracy on the lens light model. In other words $\Lambda_i$ is the main parameter that determines the maximum allowed $\alpha_i$ to have  $\delta C_{gi}$ below a desired threshold.

Equation \ref{eq:alphasimp} only applies to cases in which the lens residuals in one band are negligible.
In the general case, which is relevant for the colours other than $g-i$ in our sample, it is necessary to  use Equation \ref{eq:req}, which places joint constraints on the $\alpha_{\lambda}$s and $\Lambda_{\lambda}$s in two bands. In this case, Equation \ref{eq:req} cannot be simplified further since $\alpha_i$ and $\alpha_{\textnormal{band}}$ are not necessarily correlated. This is so because of the colour gradients in the lens light residuals, and is shown in Figure \ref{agai}, where $\alpha_g$ is plotted against $\alpha_i$.

\subsection{A proxy for the lens light contamination}
\label{sect:contrast}

\begin{figure}
    \centering
    \includegraphics[width=9cm]{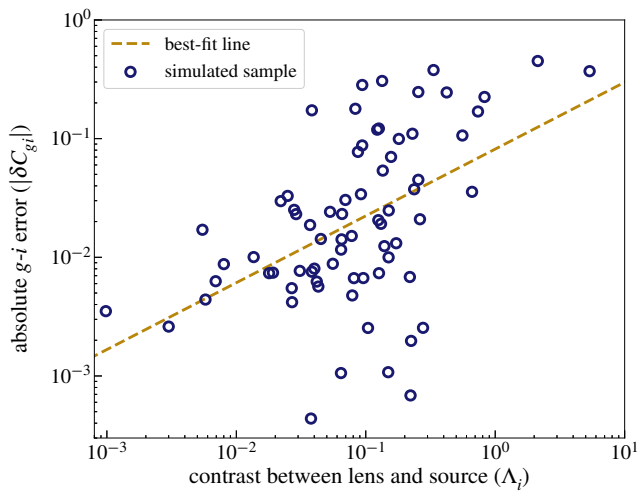}
    \caption{Source colour measurement error $|\delta C_{gi}|$ plotted against the $i$-band contrast between the lens and source  ($\Lambda_i$). The dashed line shows the best-fit power law, with a $0.563 \pm 0.014$ slope. For real strongly lensed systems this correlation can be used to select a $\Lambda_i$ threshold below which the accuracy of source colour measurement is sufficient for a given application (e.g. source photo-$z$ estimation).}
    \label{lens at source ratio}
\end{figure}

As mentioned in Section \ref{section: the accuracy of the lens light model}, $\alpha_i$ is challenging to measure for real strongly lensed systems. 
Therefore, Equations \ref{eq:req} and \ref{eq:alphasimp} cannot be used to evaluate the accuracy of the best-fit colours (and consequently, the accuracy of the estimated photo-$z$s) for real strongly lensed galaxies. 
However, the contrast between the lens and source ($\Lambda$; see Section \ref{section: the accuracy of the lens light model}) depends only on the best-fit lens and the best-fit source light models, making it straightforward to measure for real strongly lensed systems. As indicated in Equations \ref{eq:req} and \ref{eq:alphasimp}, although $\Lambda_i$ is not the only parameter determining the $\delta C_{gi}$, we expect the $\delta C_{gi}$ to be correlated with $\Lambda_i$. Hence, $\Lambda_i$ can be used to evaluate the accuracy of the inferred colours of real strongly lensed galaxies. 
Here we investigate the correlation between $\delta C_{gi}$ and $\Lambda_i$ on the best-fit models.

Figure \ref{lens at source ratio} shows  the correlation between $\delta C_{gi}$ and $\Lambda_i$. The slope of the best-fit power law (dashed line) is $0.563 \pm 0.014$. The $1\sigma$ scatter of the $\log_{10}\delta C_{gi}$ around the best-fit $\log_{10}\delta C_{gi}(\Lambda_i)$ is 0.508. 
This correlation is much weaker than the correlation between $\delta C_{gi}$ and the deviations of the lens light from a S\'{e}rsic profile or the lens residuals (e.g. see Section \ref{section: the accuracy of the lens light model}, and Figures \ref{deltaC vs deltaC plot}, \ref{alpha vs lambda plot}, and \ref{residuals}), and cannot be used to estimate the $\delta C_{gi}$ with high precision for individual systems. However, we can use this correlation to select a $\Lambda_i$ threshold, below which the accuracy of source colour measurement is sufficient for a given application (e.g. source photo-$z$ estimation). This is  shown  in Section \ref{results: real sample}, where we estimate the source photo-$z$s for our real lens sample.

\section{Results: Photo-$z$ estimation}
\label{results section}
\subsection{Simulated sample}
\label{results: simulated sample}

\begin{figure}
    \centering
    \includegraphics[width=9cm]{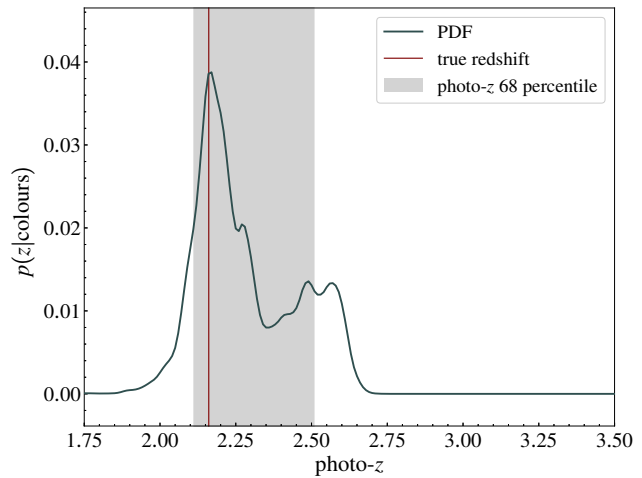}
    \caption{Example of a normalised photo-$z$ probability distribution function (PDF), obtained by running a photo-$z$ algorithm (BPZ is used here) on the best-fit source colours and their uncertainties. The red line shows the true redshift, and the grey region shows the 68th percentile of the redshift PDF. The redshift with the highest probability is used as the estimated photo-$z$.}
    \label{PDF: simulations}
\end{figure}

In this section we use the BPZ algorithm to estimate the source photo-$z$s of our simulated sample from the best-fit source magnitudes measured in Section \ref{colour measurement}.
We showed in Section \ref{colour measurement} that the colour measurement is significantly more accurate than the magnitude measurement. In other words, magnitude uncertainties are strongly correlated.
The correct way to take these correlations into account would be to provide BPZ with the $5\times5$ covariance matrix of our magnitude measurements. This feature, however, is not supported in BPZ.

Hence, we instead ran the BPZ algorithm with the source colours and their uncertainties as follows. We selected the $i$-band as the reference band of the colour measurement by assuming a small $i$-band uncertainty (0.01), and used the $1\sigma$ scatter of the colour measurement errors (measured in Section \ref{colour measurement}) as the uncertainty on the non-reference bands:  0.03, 0.02, 0.03, and 0.06 on the $g$-, $r$-, $z$-, and $y$-bands, respectively. 

\begin{figure}
    \centering
    \includegraphics[width=9cm]{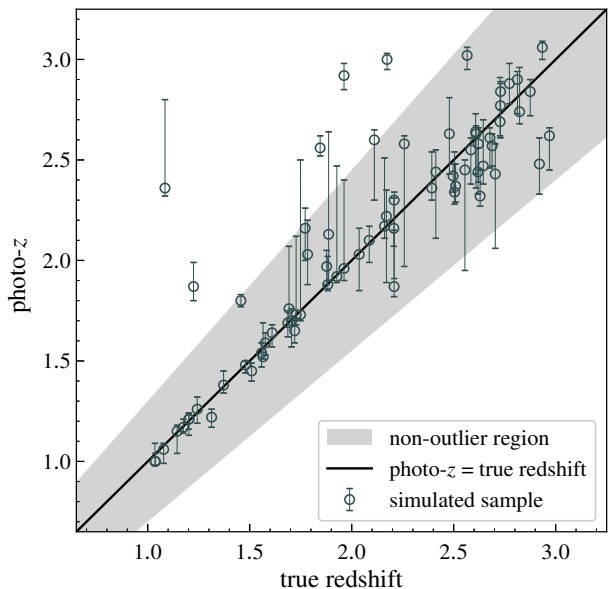} 
    \caption{Estimated photo-$z$s vs   true redshifts for the source galaxies in our simulated sample. The photo-$z$s were obtained by running the BPZ algorithm on the best-fit colours shown in Figure \ref{magnitudes and colors for fits: simulations}. The black line shows where the photo-$z$ equals the true redshift. The grey shaded region indicates where $\delta z/(1+z_{\mathrm{true}}) < 0.15$ (i.e. where the estimated redshift is not an outlier). The outlier fraction (fraction of the cases with $\delta z/(1+z_{\mathrm{true}}) > 0.15$) is 8.3$\%$ and the $1\sigma$ scatter of redshift errors ($\delta z/1+z_{\mathrm{true}}$) is 0.032. As in  Figure \ref{PDF: simulations}, the redshift with the highest probability in the redshift PDF was used as the estimated photo-$z$. The error bars in this figure correspond to the 68th percentile of the redshift PDFs (grey shaded region in Figure \ref{PDF: simulations}.)}
    \label{photoz: simulations}
\end{figure}

Figure \ref{PDF: simulations} shows an example of a normalised photo-$z$ PDF, obtained by running the BPZ on a set of best-fit source colours (from Section \ref{colour measurement}) and uncertainties. We define the estimated photo-$z$ as the redshift with the highest probability in this PDF. Figure \ref{photoz: simulations} shows the estimated photo-$z$s for all the source galaxies in our simulated sample. The error bars show the 68\% credible region of the redshift PDFs (corresponding to the shaded area in Figure \ref{PDF: simulations}). The outlier fraction for the estimated photo-$z$s, which we define as the fraction of cases with $\delta z/(1+z_{\mathrm{true}}) > 0.15$, is 8.3$\%$, and the $1\sigma$ scatter of the redshift errors ($\delta z/(1+z_{\mathrm{true}})$) is 0.032.

As discussed in Section \ref{results: propagation of modelling uncertainties into colours}, blending with the lens light is the main factor that complicates the source colour measurement, and consequently the photo-$z$ estimation. We evaluated the impact of blending on the accuracy of estimated source photo-$z$s by testing our algorithm on a sample of simulated isolated galaxies consisting of all the unlensed sources from our simulated strongly lensed sample. 
We fitted these simulated isolated galaxies with S\'{e}rsic light profiles to measure their colours, and ran the BPZ algorithm on the measured colours.
We found that there are no outliers amongst the estimated photo-$z$s, and the $1\sigma$ scatter of the redshift errors ($\delta z/(1+z_{\mathrm{true}})$) is 0.018 (Figure \ref{photoz: isolated}). 
The details of these simulations, method, and results are available in Appendix \ref{app: isolated galaxies}.
This highly accurate photo-$z$ estimation of simulated isolated galaxies implies that for simulated systems, where source colours are drawn from the SED templates of BPZ (see Section \ref{simulations}), blending (and not the performance of the photo-$z$ algorithm) is the most important source of uncertainty in photo-$z$ estimation; we discuss this in more detail in Section \ref{real sample vs the simulated sample}.

\begin{figure}
    \centering
    \includegraphics[width=9cm]{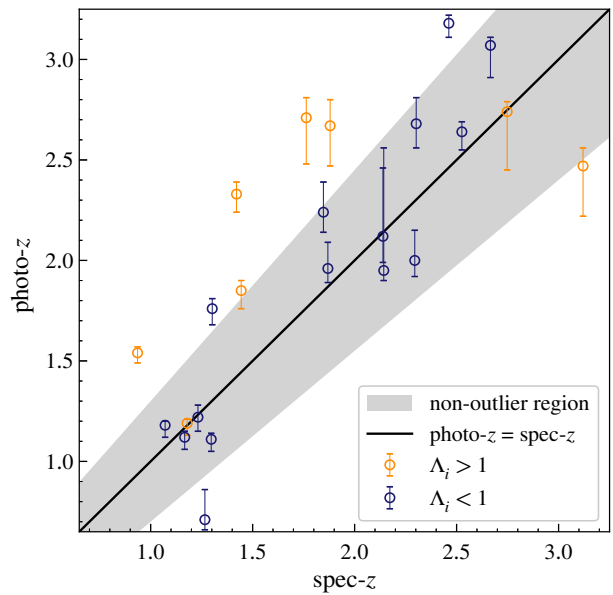}
    \caption{Estimated photo-$z$s vs the spectroscopic redshifts for the source galaxies in our sample of real strongly lensed galaxy-galaxy systems. The error bars show the $1\sigma$ uncertainty on the estimated redshift, and the grey shaded region indicates where the estimated redshift is not an outlier. The outlier fraction for the 15 systems with $\Lambda_i<1$ (blue points) is $20\%$, while the outlier fraction for the remaining 8 systems with $\Lambda_i>1$ (yellow points) is $75\%$. The source photo-$z$ estimation is much more accurate for the systems with $\Lambda_i<1$, due to their significantly more accurate source colour measurements (compared to the systems with $\Lambda_i>1$).}
    \label{photoz: real data}
\end{figure}

\subsection{Real sample}
\label{results: real sample}

In this section we estimate the source photo-$z$s for our sample of real strongly lensed galaxies. We measured the source magnitudes for our 23 real strongly lensed systems (Section \ref{real sample}) by following the approach described in  Section \ref{colour measurement}. Here, unlike Section \ref{colour measurement}, visual inspection is our only means for identifying the contaminating objects missed by {\sc SExtractor}. We identified six systems with one {\tt SExtractor}-missed contaminant and one system with two {\tt SExtractor}-missed contaminants, and modelled these contaminants with additional S\'{e}rsic light components. 

We estimated the source photo-$z$s by running BPZ on the best-fit source colours. As in Section \ref{results: simulated sample}, we chose the $i$-band as the reference band by assuming a small $i$-band uncertainty (0.01). The uncertainty in the remaining bands consists of the uncertainties from the photometric zero-points and the uncertainties from the best-fit models. The zero-point uncertainties of the HSC filters are available from  Table 6 in \cite{2018PASJ...70S...8A} (see also Section \ref{HSC photometry}). We used the colour measurement uncertainty estimates that were obtained by running our deblending algorithm on the simulated sample (Section \ref{colour measurement}) as the uncertainties from the best-fit models. These independent uncertainties add up to a total 0.04, 0.03, 0.04, and 0.06 uncertainty respectively for the $g$, $r$, $z$, and $y$ filters. 
Figure \ref{photoz: real data} shows the estimated source photo-$z$s and the 68\% credible region of their PDFs plotted against the spectroscopic redshifts. The outlier fraction (fraction of cases with $\delta z/(1+z_{\mathrm{spec}}) > 0.15$) is $39\%$ and the $1\sigma$ scatter of redshift errors ($\delta z/(1+z_{\mathrm{spec}})$) is 0.157. 
The performance is thus much worse than that found for the simulated sample.

The source photo-$z$ estimation inaccuracies are driven by the source colour measurement inaccuracies and the performance of the BPZ algorithm. We estimated the contribution of the latter by comparing the performance of our source photo-$z$ estimation (on real strongly lensed galaxies) with that of BPZ on isolated galaxies in a similar $1 < z < 3$ redshift range. We did  this by selecting a large sample ($\sim 2000$) of such galaxies in the HSC PDR2 $grizy$ photometry with spectroscopic redshifts available from The VIMOS VLT Deep Survey \citep[VVDS,][]{2013A&A...559A..14L} and 3D-HST \citep{2014ApJS..214...24S, 2016ApJS..225...27M}. Running BPZ on the CModel-measured $grizy$ magnitudes results in a $20\%$ to $30\%$ photo-$z$ outlier fraction. 

On the one hand, this result suggests that the higher outlier fraction in the photo-$z$s of the real strongly lensed sources, compared to the simulation case, is due to limitations of BPZ when dealing with relatively faint sources at $z>1$.
On the other hand, the performance is still significantly worse for the strongly lensed sources compared to the isolated galaxy case.
Hence, systematic uncertainties of source colour measurement propagate significantly into source photo-$z$ estimation, and are an important driver of the higher photo-$z$ outlier fraction of real lensed galaxies. 

We show in Section \ref{sect:contrast} that the source colour measurement errors are correlated with $\Lambda_i$, and suggested that a $\Lambda_i$ threshold can be used to identify the  high-quality colour measurements before estimating the photo-$z$s. Here, we choose $\Lambda_i = 1$ as the threshold, and investigate whether it improves the source photo-$z$ estimation;  $\Lambda_i = 1$ is the limit above which the  lens becomes brighter than the images at the image positions.

Figure \ref{lens at source ratio} shows that $\Lambda_i = 1$ roughly corresponds to $\delta C_{gi} = 0.1$. In order to estimate the expected photo-$z$ outlier fraction corresponding to this scatter on $\delta C_{gi}$, we conducted the following experiment.
Following the approach of Section \ref{simulations}, we drew a large sample (10000) of $grizy$ magnitudes from the BPZ SED templates, with $1 < z < 3$. We added Gaussian scatters to the drawn magnitudes and then used the BPZ algorithm to estimate the photo-$z$s. We found   that adding a 0.1 mag Gaussian scatter to each drawn magnitude results in a $31\%$ outlier fraction and a 0.16 $1\sigma$ scatter of the redshift errors. Since the 0.1 mag scatter roughly corresponds to $\Lambda_i = 1$ (Figure \ref{lens at source ratio}), we conclude that achieving sub-$30\%$ photo-$z$ outlier fractions with our method (comparable to the performance of the BPZ on isolated galaxies in the same redshift range; see further  details in Section \ref{real sample vs the simulated sample}) requires sub-unity $\Lambda_i$.


Following the approach described in  Section \ref{sect:contrast}, we measured the $\Lambda_i$ for all the systems in our real sample from their best-fit lens and source light models. We find 15 systems with $\Lambda_i < 1$ (in  blue in Figure \ref{photoz: real data}), for which the photo-$z$ outlier fraction is $20\%$ and the $1\sigma$ scatter of redshift errors is 0.089. The remaining eight systems with $\Lambda_i > 1$ (in yellow in Figure \ref{photoz: real data}) have a $75\%$ photo-$z$ outlier fraction and a 0.312 $1\sigma$ scatter of the redshift errors.

\section{Discussion}
\label{discussion}

\subsection{Real sample versus the simulated sample}
\label{real sample vs the simulated sample}

The source photo-$z$ estimation of the systems in the real sample is significantly less accurate than that of the systems in the simulated sample ($39\%$ vs $8.3\%$ photo-$z$ outlier fraction). Even after imposing the $\Lambda_i < 1$ threshold, the photo-$z$ outlier fraction of the real sample ($20\%$) remains significantly larger than that of the simulated sample. 
The uncertainties in modelling the lens light are not expected to propagate significantly into the photo-$z$ estimation of the real systems with $\Lambda_i < 1$ (see Sections \ref{sect:contrast} and \ref{results: real sample}). Since the uncertainties on modelling the lens light are the main drivers of the uncertainties on source colour measurement (see Section \ref{section: the accuracy of the lens light model}), the less-accurate photo-$z$ estimation of these systems can mainly be attributed to the performance of the BPZ algorithm.

It is straightforward for BPZ to fit the source colours of the simulated sample because these colours have been drawn from its own SED templates. 
This is evident from our experiment at the end of Section \ref{results: simulated sample} (detailed in Appendix \ref{app: isolated galaxies}); in the absence of lensing there are no photo-$z$ outliers for simulated isolated galaxies in the $1<z<3$ range. This indicates that even the small $8.3\%$ outlier fraction of the photo-$z$ estimation of simulated strongly lensed galaxies in Section \ref{results: simulated sample} was mainly caused by lensing.
In contrast, real colours are much more challenging to fit by BPZ, and a $20-30\%$ photo-$z$ outlier fraction is typical for galaxies in $1<z<3$ with $grizy$ photometry of the HSC (see Section \ref{results: real sample}). This is partly due to the template incompleteness of the BPZ at this redshift range. Moreover, the performance of the BPZ and similar SED-fitting algorithms drop when the $4000\textnormal{\r{A}}$ break gets redshifted out of the reddest filter. This occurs because the $4000\textnormal{\r{A}}$ break is one of the primary spectral features used for fitting the spectra of galaxies in SED-fitting photo-$z$ algorithms. The $4000\textnormal{\r{A}}$ break gets shifted redwards of the $y$-band (the reddest filter of HSC, $9300-10700\textnormal{\r{A}}$) at $z\sim1.75$. As can be seen for the $z>1.75$ galaxies in Figure \ref{photoz: simulations}, this effect  even propagates noticeably into the photo-$z$ estimation of the simulated sources. The template incompleteness exacerbates these effects for the real galaxies at $z > 1.75$.

The higher photo-$z$ outlier fraction of the real sample could also be caused by complications in modelling the surface brightness of real sources compared to the simulated ones ; the source model is accurate by construction in the simulated sample, but not for the real sample. Additionally, simulating the lensing potential with a simple SIE model (and making it easy to fit by the SIE mass model in our deblending algorithm) contributes to the difference between the photo-$z$ outlier fractions of the simulated and the real samples, although to a much lesser degree than the source light modelling (since lensing is achromatic).

Nevertheless, the inaccuracies of lens light modelling govern the source colour measurement inaccuracy. In Section \ref{results: propagation of modelling uncertainties into colours}, we showed that this is the case for our simulated lensed systems. We can argue that this is also the case for the real lensed systems; the estimated source photo-$z$s are dramatically more accurate for the real systems with $\Lambda_i<1$ (for which the lens light modelling inaccuracies are not expected to propagate significantly into photo-$z$ estimation) compared to the real systems with $\Lambda_i>1$ ($20\%$ vs $75\%$ photo-$z$ outlier fraction). Moreover, the accuracy of the estimated photo-$z$s for the real sources with $\Lambda_i<1$ is consistent with the expected accuracy of BPZ for real galaxies in a similar redshift range ($20-30\%$ photo-$z$ outlier fraction, if the $grizy$ photometry of the HSC is used). Hence, other systematics (such as modelling the light of the real source with a S\'{e}rsic profile or modelling the lensing potential with an SIE) propagate into colour measurement to a much lesser degree than the lens light modelling inaccuracies.

Stage IV surveys are expected to improve the accuracy of estimated source photo-$z$s, thanks to their better-calibrated photometry and broader spectral coverage.
In this work the relative zero-point calibration accuracy of HSC photometry sets the lower limit on the accuracy of source colour measurement (see Sections \ref{HSC photometry}, \ref{section: the accuracy of the lens light model}, \ref{results: real sample}, and \ref{lens light modelling innacuracy}). 
The better-calibrated photometry of Euclid and the LSST will lower this threshold, making the uncertainties from lens light modelling even more dominant than their current level with HSC photometry. 
Moreover, at $z > 1.7$, stage IV surveys (Euclid in particular) will enable photo-$z$ estimations (for lensed sources as well as isolated galaxies) that are significantly more accurate than the HSC-based measurements. 
This is because stage IV surveys will provide photometry in near-infrared filters redder than the $y$-band of HSC, which will be instrumental in breaking the redshift degeneracies by detecting the $4000\textnormal{\r{A}}$ break.
Unless the contamination from the lens galaxy is properly dealt with, however, these improvements will only translate into more accurate photo-$z$s for the systems with a low $\Lambda_i$.
This emphasises the importance of improving the deblending algorithms (lens light modelling in particular) in order to estimate accurate source photo-$z$s for large samples of lenses.

\subsection{Lens light modelling}
\label{lens light modelling innacuracy}

Since the uncertainties from the lens light model dominate the source colour measurement uncertainties (see Section \ref{real sample vs the simulated sample}), the accuracy of the source colour measurement and photo-$z$ estimation can be estimated for any sample of real strongly lensed galaxies based on their $\Lambda_i$ distributions. The large scatter on the best-fit line in Figure \ref{lens at source ratio} prevents high-precision estimations of the accuracy of measured colours and estimated photo-$z$s for individual systems. However, systems with accurately measurable colours and photo-$z$s can be identified by imposing a $\Lambda_i$ threshold, as shown in Section \ref{results: real sample}. Although it dramatically increases the accuracy of photo-$z$ estimation, this practice hinders population studies that require complete and representative lens samples (e.g. for studying the inner structure of lens galaxies) by introducing non-negligible selection biases (e.g. towards the lens galaxies that are fainter or have a larger Einstein radius). Although followed-up source spectroscopic redshifts can compensate for the lack of accurate photo-$z$s at high $\Lambda_i$, follow-up spectroscopy campaigns can be carried out only for a small fraction of the large samples expected to be discovered with the LSST and Euclid. Hence, without accurate photo-$z$s at high $\Lambda_i$, our ability to take full advantage of these large samples remains limited. 

Therefore, it is essential to improve this and similar methods by improving the deblending and photo-$z$ algorithms. Improvements in the photo-$z$ algorithm result in an overall increase in the accuracy of the estimated photo-$z$s. However, these improvements do not translate into increased source photo-$z$ accuracy for the systems with a large $\Lambda_i$: inaccurate colours can hardly yield accurate photo-$z$s. Hence, improving the deblending algorithm is a necessity for systems with a high $\Lambda_i$.  

For samples of lenses with similar properties to those considered in this work, modelling the lens light with a single S\'{e}rsic profile is the main source of colour measurement error. More complex lens light models, such as multi-component or non-parametrised descriptions of the surface brightness distribution, can potentially yield smaller lens light residuals (i.e. lower $\alpha_i$) and a more accurate source colour measurement (see Section \ref{section: the accuracy of the lens light model}, Appendix \ref{sect:lensres}, and Figures \ref{alpha vs lambda plot} and \ref{residuals} for the correlation between $\delta C_{gi}$ and residuals).

\subsection{Contaminating bright objects}

One of the potential complications arising in applying the methods of this paper to estimate the source photo-$z$s of large samples of strong lenses (more than a few hundred systems) is the identification of contaminating bright objects. Contaminating objects that are faint and not closely aligned with the lensed system can be automatically identified and masked with segmenting algorithms such as {\sc SExtractor}. However, these algorithms cannot distinguish the bright contaminants that are closely aligned with the lensed system from the lensed system itself. These missed contaminants, if unidentified (and not modelled), significantly increase the source colour measurement (and photo-$z$ estimation) errors. 

In this work visual inspection has been our only means for identifying the {\tt SExtractor}-missed contaminating objects. However, repeating the same practice for the large upcoming samples is inefficient. Currently, this is an  open problem for the analysis of large samples of lenses, and one that affects not only the measurement of the source photo-$z$, but more generally the lens modelling process.

\subsection{Prior on the source properties}

Inferring the redshift of a galaxy requires making prior assumptions on its probability distribution. In a Bayesian approach, such as the one adopted in this paper, the prior is explicitly included in the formalism. In methods based on machine learning, the choice of the distribution of the training sample acts as a prior on the inference.
In our experiment we assumed a flat prior in redshift-magnitude space. The rate of catastrophic failures in photo-$z$ estimates, quantified by the outlier fraction, is not very sensitive to this choice: we verified that the results do not change significantly when adopting the default magnitude-dependent redshift prior of BPZ.
However, assuming the wrong prior can systematically bias the inferred photo-$z$s, potentially limiting our ability to take advantage of the statistical combination of large samples of lenses.

Choosing the correct prior for a sample of strongly lensed sources is not a trivial task.
The distribution in redshift-magnitude space of lensed galaxies differs from that of isolated objects because of strong lensing selection effects: the probability of a galaxy being strongly lensed increases as a function of redshift, and the lensing magnification allows us to detect intrinsically fainter objects. For this reason, a prior calibrated on a sample of isolated galaxies is not a good choice for strongly lensed sources.
The main challenge in determining the correct prior distribution of lensed sources is  that the redshift of a source is correlated with the mass distribution of the lens, and both of these properties are, in all practical cases, unknown.
The correct way of solving this problem is to jointly describe the properties of the foreground and background galaxy population with a hierarchical model, following the method proposed by \citet{Son22b}.
This approach allows the mass distribution of the lenses and the redshift distribution of the sources to be inferred simultaneously; however, it requires the lensing selection effects to be explicitly accounted for.

\section{Conclusion}
\label{conclusion}

We showed that the accuracy of the lens light model and the contrast between the lens and source are the main drivers of the uncertainties on source colour measurement and photo-$z$ estimation. We modelled the galaxy-galaxy strong lens systems by simultaneously modelling the lensing potential with an SIE mass profile, the lens surface brightness with a S\'{e}rsic profile, and the source surface brightness with a S\'{e}rsic profile that is lensed through the lensing potential. If the lens is not brighter than the source at the source image position ($\Lambda_i < 1$), this modelling step provides source colours that are accurate enough to render the performance of the photo-$z$ algorithm the main driver of the uncertainties on source photo-$z$ estimation.

If the lens becomes brighter than the source at the source image position ($\Lambda_i > 1$), the propagation of the lens residuals at the source image position into the best-fit source light model reduces the ability of our deblending algorithm in measuring source colours that are sufficiently accurate for an accurate source photo-$z$ estimation (i.e. $\delta C_{gi}$ becomes larger than 0.1). In this scenario, the uncertainties in modelling the lens light become the dominant source of the uncertainties on source photo-$z$ estimation. 

We showed that the maximum allowed deviation of the lens light from a S\'{e}rsic profile (i.e. the maximum allowed $\alpha_i$) that is required to measure source colours that are more accurate than a given threshold, decreases with increasing $\Lambda_i$ (Section \ref{section: the accuracy of the lens light model} and Figure \ref{alpha vs lambda plot}). Hence, $\delta C_{gi}$ increases with increasing $\Lambda_i$. We inferred this correlation from a simulated sample of strongly lensed systems. Despite the large scatter on this relation, it can be used to filter the objects with accurately measurable source colours.

However, filtering introduces non-negligible selection biases towards lenses that are fainter or have a larger Einstein radius. This in turn hinders our ability to take full advantage of the current and upcoming large samples of strongly lensed galaxy-galaxy systems (from the ESA-Euclid, the Rubin-LSST, and the CSST) in studying complete and representative samples of lens galaxies \citep[which are needed to map the properties of the lens population to those of the general galaxy population, as shown by][]{Son22a}. Obtaining complete lens samples requires improvements in both the photo-$z$ and the deblending algorithms. In particular, more advanced lens light models (compared to the single S\'{e}rsic profile used here) that leave behind smaller residuals will raise the $\Lambda_i$ threshold below which source colours can be measured accurately. 
In the case that obtaining accurate photo-$z$s of complete samples of strong lenses  proves difficult, an alternative approach is to use  information on the population distribution of galaxy redshifts, as opposed to individual source redshift measurements, and to incorporate it into a Bayesian hierarchical model that takes lensing selection effects into account. This approach is discussed in detail by \citet{Son22b}.

Significant improvements in both the deblending and photo-$z$ algorithms remain essential before the launch of Euclid and the LSST. Nonetheless, this work is a first exploratory step in constructing the pipeline for estimating the source redshifts of the upcoming samples of strong lenses.

\begin{acknowledgement}

DL would like to thank Jens Hjorth for constructive conversations, which helped improve the results and shaped parts of the discussion. 
The authors would like to thank the anonymous referee for constructive comments.
DL is supported by a grant from VILLUM FONDEN (project number 16599). 
AA’s research is funded by Villum Experiment Grant \textit{Cosmic Beacons} (project number 36225).

\end{acknowledgement}

%
%

\bibliography{aanda}

\begin{thebibliography}{75}
\expandafter\ifx\csname natexlab\endcsname\relax\def\natexlab#1{#1}\fi

\bibitem[{{Ahn} {et~al.}(2012){Ahn}, {Alexandroff}, {Allende Prieto},
  {Anderson}, {Anderton}, {Andrews}, {Aubourg}, {Bailey}, {Balbinot}, {Barnes},
  {Bautista}, {Beers}, {Beifiori}, {Berlind}, {Bhardwaj}, {Bizyaev}, {Blake},
  {Blanton}, {Blomqvist}, {Bochanski}, {Bolton}, {Borde}, {Bovy}, {Brandt},
  {Brinkmann}, {Brown}, {Brownstein}, {Bundy}, {Busca}, {Carithers}, {Carnero},
  {Carr}, {Casetti-Dinescu}, {Chen}, {Chiappini}, {Comparat}, {Connolly},
  {Crepp}, {Cristiani}, {Croft}, {Cuesta}, {da Costa}, {Davenport}, {Dawson},
  {de Putter}, {De Lee}, {Delubac}, {Dhital}, {Ealet}, {Ebelke}, {Edmondson},
  {Eisenstein}, {Escoffier}, {Esposito}, {Evans}, {Fan}, {Femen{\'\i}a
  Castell{\'a}}, {Fern{\'a}ndez Alvar}, {Ferreira}, {Filiz Ak}, {Finley},
  {Fleming}, {Font-Ribera}, {Frinchaboy}, {Garc{\'\i}a-Hern{\'a}ndez},
  {Garc{\'\i}a P{\'e}rez}, {Ge}, {G{\'e}nova-Santos}, {Gillespie}, {Girardi},
  {Gonz{\'a}lez Hern{\'a}ndez}, {Grebel}, {Gunn}, {Guo}, {Haggard}, {Hamilton},
  {Harris}, {Hawley}, {Hearty}, {Ho}, {Hogg}, {Holtzman}, {Honscheid},
  {Huehnerhoff}, {Ivans}, {Ivezi{\'c}}, {Jacobson}, {Jiang}, {Johansson},
  {Johnson}, {Kauffmann}, {Kirkby}, {Kirkpatrick}, {Klaene}, {Knapp}, {Kneib},
  {Le Goff}, {Leauthaud}, {Lee}, {Lee}, {Long}, {Loomis}, {Lucatello},
  {Lundgren}, {Lupton}, {Ma}, {Ma}, {MacDonald}, {Mack}, {Mahadevan}, {Maia},
  {Majewski}, {Makler}, {Malanushenko}, {Malanushenko}, {Manchado},
  {Mandelbaum}, {Manera}, {Maraston}, {Margala}, {Martell}, {McBride},
  {McGreer}, {McMahon}, {M{\'e}nard}, {Meszaros}, {Miralda-Escud{\'e}},
  {Montero-Dorta}, {Montesano}, {Morrison}, {Muna}, {Munn}, {Murayama},
  {Myers}, {Neto}, {Nguyen}, {Nichol}, {Nidever}, {Noterdaeme}, {Nuza},
  {Ogando}, {Olmstead}, {Oravetz}, {Owen}, {Padmanabhan},
  {Palanque-Delabrouille}, {Pan}, {Parejko}, {Parihar}, {P{\^a}ris},
  {Pattarakijwanich}, {Pepper}, {Percival}, {P{\'e}rez-Fournon},
  {P{\'e}rez-R{\`a}fols}, {Petitjean}, {Pforr}, {Pieri}, {Pinsonneault}, {Porto
  de Mello}, {Prada}, {Price-Whelan}, {Raddick}, {Rebolo}, {Rich}, {Richards},
  {Robin}, {Rocha-Pinto}, {Rockosi}, {Roe}, {Ross}, {Ross}, {Rossi},
  {Rubi{\~n}o-Martin}, {Samushia}, {Sanchez Almeida}, {S{\'a}nchez},
  {Santiago}, {Sayres}, {Schlegel}, {Schlesinger}, {Schmidt}, {Schneider},
  {Schultheis}, {Schwope}, {Sc{\'o}ccola}, {Seljak}, {Sheldon}, {Shen}, {Shu},
  {Simmerer}, {Simmons}, {Skibba}, {Skrutskie}, {Slosar}, {Sobreira}, {Sobeck},
  {Stassun}, {Steele}, {Steinmetz}, {Strauss}, {Streblyanska}, {Suzuki},
  {Swanson}, {Tal}, {Thakar}, {Thomas}, {Thompson}, {Tinker}, {Tojeiro},
  {Tremonti}, {Vargas Maga{\~n}a}, {Verde}, {Viel}, {Vikas}, {Vogt}, {Wake},
  {Wang}, {Weaver}, {Weinberg}, {Weiner}, {West}, {White}, {Wilson},
  {Wisniewski}, {Wood-Vasey}, {Yanny}, {Y{\`e}che}, {York}, {Zamora},
  {Zasowski}, {Zehavi}, {Zhao}, {Zheng}, {Zhu}, \&
  {Zinn}}]{2012ApJS..203...21A}
{Ahn}, C.~P., {Alexandroff}, R., {Allende Prieto}, C., {et~al.} 2012, \apjs,
  203, 21

\bibitem[{{Aihara} {et~al.}(2019){Aihara}, {AlSayyad}, {Ando}, {Armstrong},
  {Bosch}, {Egami}, {Furusawa}, {Furusawa}, {Goulding}, {Harikane}, {Hikage},
  {Ho}, {Hsieh}, {Huang}, {Ikeda}, {Imanishi}, {Ito}, {Iwata}, {Jaelani},
  {Kakuma}, {Kawana}, {Kikuta}, {Kobayashi}, {Koike}, {Komiyama}, {Li},
  {Liang}, {Lin}, {Luo}, {Lupton}, {Lust}, {MacArthur}, {Matsuoka}, {Mineo},
  {Miyatake}, {Miyazaki}, {More}, {Murata}, {Namiki}, {Nishizawa}, {Oguri},
  {Okabe}, {Okamoto}, {Okura}, {Ono}, {Onodera}, {Onoue}, {Osato}, {Ouchi},
  {Shibuya}, {Strauss}, {Sugiyama}, {Suto}, {Takada}, {Takagi}, {Takata},
  {Takita}, {Tanaka}, {Terai}, {Toba}, {Uchiyama}, {Utsumi}, {Wang}, {Wang}, \&
  {Yamada}}]{2019PASJ...71..114A}
{Aihara}, H., {AlSayyad}, Y., {Ando}, M., {et~al.} 2019, \pasj, 71, 114

\bibitem[{{Aihara} {et~al.}(2018{\natexlab{a}}){Aihara}, {Arimoto},
  {Armstrong}, {Arnouts}, {Bahcall}, {Bickerton}, {Bosch}, {Bundy}, {Capak},
  {Chan}, {Chiba}, {Coupon}, {Egami}, {Enoki}, {Finet}, {Fujimori}, {Fujimoto},
  {Furusawa}, {Furusawa}, {Goto}, {Goulding}, {Greco}, {Greene}, {Gunn},
  {Hamana}, {Harikane}, {Hashimoto}, {Hattori}, {Hayashi}, {Hayashi},
  {He{\l}miniak}, {Higuchi}, {Hikage}, {Ho}, {Hsieh}, {Huang}, {Huang},
  {Ikeda}, {Imanishi}, {Inoue}, {Iwasawa}, {Iwata}, {Jaelani}, {Jian},
  {Kamata}, {Karoji}, {Kashikawa}, {Katayama}, {Kawanomoto}, {Kayo}, {Koda},
  {Koike}, {Kojima}, {Komiyama}, {Konno}, {Koshida}, {Koyama}, {Kusakabe},
  {Leauthaud}, {Lee}, {Lin}, {Lin}, {Lupton}, {Mandelbaum}, {Matsuoka},
  {Medezinski}, {Mineo}, {Miyama}, {Miyatake}, {Miyazaki}, {Momose}, {More},
  {More}, {Moritani}, {Moriya}, {Morokuma}, {Mukae}, {Murata}, {Murayama},
  {Nagao}, {Nakata}, {Niida}, {Niikura}, {Nishizawa}, {Obuchi}, {Oguri},
  {Oishi}, {Okabe}, {Okamoto}, {Okura}, {Ono}, {Onodera}, {Onoue}, {Osato},
  {Ouchi}, {Price}, {Pyo}, {Sako}, {Sawicki}, {Shibuya}, {Shimasaku},
  {Shimono}, {Shirasaki}, {Silverman}, {Simet}, {Speagle}, {Spergel},
  {Strauss}, {Sugahara}, {Sugiyama}, {Suto}, {Suyu}, {Suzuki}, {Tait},
  {Takada}, {Takata}, {Tamura}, {Tanaka}, {Tanaka}, {Tanaka}, {Tanaka},
  {Terai}, {Terashima}, {Toba}, {Tominaga}, {Toshikawa}, {Turner}, {Uchida},
  {Uchiyama}, {Umetsu}, {Uraguchi}, {Urata}, {Usuda}, {Utsumi}, {Wang}, {Wang},
  {Wong}, {Yabe}, {Yamada}, {Yamanoi}, {Yasuda}, {Yeh}, {Yonehara}, \&
  {Yuma}}]{Aih++18a}
{Aihara}, H., {Arimoto}, N., {Armstrong}, R., {et~al.} 2018{\natexlab{a}},
  \pasj, 70, S4

\bibitem[{{Aihara} {et~al.}(2018{\natexlab{b}}){Aihara}, {Armstrong},
  {Bickerton}, {Bosch}, {Coupon}, {Furusawa}, {Hayashi}, {Ikeda}, {Kamata},
  {Karoji}, {Kawanomoto}, {Koike}, {Komiyama}, {Lang}, {Lupton}, {Mineo},
  {Miyatake}, {Miyazaki}, {Morokuma}, {Obuchi}, {Oishi}, {Okura}, {Price},
  {Takata}, {Tanaka}, {Tanaka}, {Tanaka}, {Uchida}, {Uraguchi}, {Utsumi},
  {Wang}, {Yamada}, {Yamanoi}, {Yasuda}, {Arimoto}, {Chiba}, {Finet},
  {Fujimori}, {Fujimoto}, {Furusawa}, {Goto}, {Goulding}, {Gunn}, {Harikane},
  {Hattori}, {Hayashi}, {He{\l}miniak}, {Higuchi}, {Hikage}, {Ho}, {Hsieh},
  {Huang}, {Huang}, {Imanishi}, {Iwata}, {Jaelani}, {Jian}, {Kashikawa},
  {Katayama}, {Kojima}, {Konno}, {Koshida}, {Kusakabe}, {Leauthaud}, {Lee},
  {Lin}, {Lin}, {Mandelbaum}, {Matsuoka}, {Medezinski}, {Miyama}, {Momose},
  {More}, {More}, {Mukae}, {Murata}, {Murayama}, {Nagao}, {Nakata}, {Niida},
  {Niikura}, {Nishizawa}, {Oguri}, {Okabe}, {Ono}, {Onodera}, {Onoue}, {Ouchi},
  {Pyo}, {Shibuya}, {Shimasaku}, {Simet}, {Speagle}, {Spergel}, {Strauss},
  {Sugahara}, {Sugiyama}, {Suto}, {Suzuki}, {Tait}, {Takada}, {Terai}, {Toba},
  {Turner}, {Uchiyama}, {Umetsu}, {Urata}, {Usuda}, {Yeh}, \&
  {Yuma}}]{2018PASJ...70S...8A}
{Aihara}, H., {Armstrong}, R., {Bickerton}, S., {et~al.} 2018{\natexlab{b}},
  \pasj, 70, S8

\bibitem[{{Auger} {et~al.}(2009){Auger}, {Treu}, {Bolton}, {Gavazzi},
  {Koopmans}, {Marshall}, {Bundy}, \& {Moustakas}}]{Aug++09}
{Auger}, M.~W., {Treu}, T., {Bolton}, A.~S., {et~al.} 2009, \apj, 705, 1099

\bibitem[{{Auger} {et~al.}(2010){Auger}, {Treu}, {Bolton}, {Gavazzi},
  {Koopmans}, {Marshall}, {Moustakas}, \& {Burles}}]{Aug++10}
{Auger}, M.~W., {Treu}, T., {Bolton}, A.~S., {et~al.} 2010, \apj, 724, 511

\bibitem[{{Barnab{\`e}} {et~al.}(2011){Barnab{\`e}}, {Czoske}, {Koopmans},
  {Treu}, \& {Bolton}}]{Bar++11}
{Barnab{\`e}}, M., {Czoske}, O., {Koopmans}, L. V.~E., {Treu}, T., \& {Bolton},
  A.~S. 2011, \mnras, 415, 2215

\bibitem[{{Ben{\'\i}tez}(2000)}]{2000ApJ...536..571B}
{Ben{\'\i}tez}, N. 2000, \apj, 536, 571

\bibitem[{{Ben{\'\i}tez} {et~al.}(2004){Ben{\'\i}tez}, {Ford}, {Bouwens},
  {Menanteau}, {Blakeslee}, {Gronwall}, {Illingworth}, {Meurer}, {Broadhurst},
  {Clampin}, {Franx}, {Hartig}, {Magee}, {Sirianni}, {Ardila}, {Bartko},
  {Brown}, {Burrows}, {Cheng}, {Cross}, {Feldman}, {Golimowski}, {Infante},
  {Kimble}, {Krist}, {Lesser}, {Levay}, {Martel}, {Miley}, {Postman}, {Rosati},
  {Sparks}, {Tran}, {Tsvetanov}, {White}, \& {Zheng}}]{2004ApJS..150....1B}
{Ben{\'\i}tez}, N., {Ford}, H., {Bouwens}, R., {et~al.} 2004, \apjs, 150, 1

\bibitem[{{Bertin} \& {Arnouts}(1996)}]{1996A&AS..117..393B}
{Bertin}, E. \& {Arnouts}, S. 1996, \aaps, 117, 393

\bibitem[{{Bolton} {et~al.}(2012){Bolton}, {Brownstein}, {Kochanek}, {Shu},
  {Schlegel}, {Eisenstein}, {Wake}, {Connolly}, {Maraston}, {Arneson}, \&
  {Weaver}}]{Bol++12}
{Bolton}, A.~S., {Brownstein}, J.~R., {Kochanek}, C.~S., {et~al.} 2012, \apj,
  757, 82

\bibitem[{{Bolton} {et~al.}(2008){Bolton}, {Burles}, {Koopmans}, {Treu},
  {Gavazzi}, {Moustakas}, {Wayth}, \& {Schlegel}}]{2008ApJ...682..964B}
{Bolton}, A.~S., {Burles}, S., {Koopmans}, L. V.~E., {et~al.} 2008, \apj, 682,
  964

\bibitem[{{Bolton} {et~al.}(2006){Bolton}, {Burles}, {Koopmans}, {Treu}, \&
  {Moustakas}}]{Bol++06}
{Bolton}, A.~S., {Burles}, S., {Koopmans}, L. V.~E., {Treu}, T., \&
  {Moustakas}, L.~A. 2006, \apj, 638, 703

\bibitem[{{Ca{\~n}ameras} {et~al.}(2021){Ca{\~n}ameras}, {Schuldt}, {Shu},
  {Suyu}, {Taubenberger}, {Meinhardt}, {Leal-Taix{\'e}}, {Chao}, {Inoue},
  {Jaelani}, \& {More}}]{Can++21}
{Ca{\~n}ameras}, R., {Schuldt}, S., {Shu}, Y., {et~al.} 2021, \aap, 653, L6

\bibitem[{{Ciotti} \& {Bertin}(1999)}]{1999A&A...352..447C}
{Ciotti}, L. \& {Bertin}, G. 1999, \aap, 352, 447

\bibitem[{{Coe} {et~al.}(2006){Coe}, {Ben{\'\i}tez}, {S{\'a}nchez}, {Jee},
  {Bouwens}, \& {Ford}}]{2006AJ....132..926C}
{Coe}, D., {Ben{\'\i}tez}, N., {S{\'a}nchez}, S.~F., {et~al.} 2006, \aj, 132,
  926

\bibitem[{{Coleman} {et~al.}(1980){Coleman}, {Wu}, \&
  {Weedman}}]{1980ApJS...43..393C}
{Coleman}, G.~D., {Wu}, C.~C., \& {Weedman}, D.~W. 1980, \apjs, 43, 393

\bibitem[{{Collett}(2015)}]{2015ApJ...811...20C}
{Collett}, T.~E. 2015, \apj, 811, 20

\bibitem[{{Dark Energy Survey Collaboration} {et~al.}(2016){Dark Energy Survey
  Collaboration}, {Abbott}, {Abdalla}, {Aleksi{\'c}}, {Allam}, {Amara},
  {Bacon}, {Balbinot}, {Banerji}, {Bechtol}, {Benoit-L{\'e}vy}, {Bernstein},
  {Bertin}, {Blazek}, {Bonnett}, {Bridle}, {Brooks}, {Brunner}, {Buckley-Geer},
  {Burke}, {Caminha}, {Capozzi}, {Carlsen}, {Carnero-Rosell}, {Carollo},
  {Carrasco-Kind}, {Carretero}, {Castander}, {Clerkin}, {Collett}, {Conselice},
  {Crocce}, {Cunha}, {D'Andrea}, {da Costa}, {Davis}, {Desai}, {Diehl},
  {Dietrich}, {Dodelson}, {Doel}, {Drlica-Wagner}, {Estrada}, {Etherington},
  {Evrard}, {Fabbri}, {Finley}, {Flaugher}, {Foley}, {Fosalba}, {Frieman},
  {Garc{\'\i}a-Bellido}, {Gaztanaga}, {Gerdes}, {Giannantonio}, {Goldstein},
  {Gruen}, {Gruendl}, {Guarnieri}, {Gutierrez}, {Hartley}, {Honscheid}, {Jain},
  {James}, {Jeltema}, {Jouvel}, {Kessler}, {King}, {Kirk}, {Kron}, {Kuehn},
  {Kuropatkin}, {Lahav}, {Li}, {Lima}, {Lin}, {Maia}, {Makler}, {Manera},
  {Maraston}, {Marshall}, {Martini}, {McMahon}, {Melchior}, {Merson}, {Miller},
  {Miquel}, {Mohr}, {Morice-Atkinson}, {Naidoo}, {Neilsen}, {Nichol}, {Nord},
  {Ogando}, {Ostrovski}, {Palmese}, {Papadopoulos}, {Peiris}, {Peoples},
  {Percival}, {Plazas}, {Reed}, {Refregier}, {Romer}, {Roodman}, {Ross},
  {Rozo}, {Rykoff}, {Sadeh}, {Sako}, {S{\'a}nchez}, {Sanchez}, {Santiago},
  {Scarpine}, {Schubnell}, {Sevilla-Noarbe}, {Sheldon}, {Smith}, {Smith},
  {Soares-Santos}, {Sobreira}, {Soumagnac}, {Suchyta}, {Sullivan}, {Swanson},
  {Tarle}, {Thaler}, {Thomas}, {Thomas}, {Tucker}, {Vieira}, {Vikram},
  {Walker}, {Wechsler}, {Weller}, {Wester}, {Whiteway}, {Wilcox}, {Yanny},
  {Zhang}, \& {Zuntz}}]{DES16}
{Dark Energy Survey Collaboration}, {Abbott}, T., {Abdalla}, F.~B., {et~al.}
  2016, \mnras, 460, 1270

\bibitem[{{Dawson} {et~al.}(2013){Dawson}, {Schlegel}, {Ahn}, {Anderson},
  {Aubourg}, {Bailey}, {Barkhouser}, {Bautista}, {Beifiori}, {Berlind},
  {Bhardwaj}, {Bizyaev}, {Blake}, {Blanton}, {Blomqvist}, {Bolton}, {Borde},
  {Bovy}, {Brandt}, {Brewington}, {Brinkmann}, {Brown}, {Brownstein}, {Bundy},
  {Busca}, {Carithers}, {Carnero}, {Carr}, {Chen}, {Comparat}, {Connolly},
  {Cope}, {Croft}, {Cuesta}, {da Costa}, {Davenport}, {Delubac}, {de Putter},
  {Dhital}, {Ealet}, {Ebelke}, {Eisenstein}, {Escoffier}, {Fan}, {Filiz Ak},
  {Finley}, {Font-Ribera}, {G{\'e}nova-Santos}, {Gunn}, {Guo}, {Haggard},
  {Hall}, {Hamilton}, {Harris}, {Harris}, {Ho}, {Hogg}, {Holder}, {Honscheid},
  {Huehnerhoff}, {Jordan}, {Jordan}, {Kauffmann}, {Kazin}, {Kirkby}, {Klaene},
  {Kneib}, {Le Goff}, {Lee}, {Long}, {Loomis}, {Lundgren}, {Lupton}, {Maia},
  {Makler}, {Malanushenko}, {Malanushenko}, {Mandelbaum}, {Manera}, {Maraston},
  {Margala}, {Masters}, {McBride}, {McDonald}, {McGreer}, {McMahon}, {Mena},
  {Miralda-Escud{\'e}}, {Montero-Dorta}, {Montesano}, {Muna}, {Myers},
  {Naugle}, {Nichol}, {Noterdaeme}, {Nuza}, {Olmstead}, {Oravetz}, {Oravetz},
  {Owen}, {Padmanabhan}, {Palanque-Delabrouille}, {Pan}, {Parejko},
  {P{\^a}ris}, {Percival}, {P{\'e}rez-Fournon}, {P{\'e}rez-R{\`a}fols},
  {Petitjean}, {Pfaffenberger}, {Pforr}, {Pieri}, {Prada}, {Price-Whelan},
  {Raddick}, {Rebolo}, {Rich}, {Richards}, {Rockosi}, {Roe}, {Ross}, {Ross},
  {Rossi}, {Rubi{\~n}o-Martin}, {Samushia}, {S{\'a}nchez}, {Sayres}, {Schmidt},
  {Schneider}, {Sc{\'o}ccola}, {Seo}, {Shelden}, {Sheldon}, {Shen}, {Shu},
  {Slosar}, {Smee}, {Snedden}, {Stauffer}, {Steele}, {Strauss}, {Streblyanska},
  {Suzuki}, {Swanson}, {Tal}, {Tanaka}, {Thomas}, {Tinker}, {Tojeiro},
  {Tremonti}, {Vargas Maga{\~n}a}, {Verde}, {Viel}, {Wake}, {Watson}, {Weaver},
  {Weinberg}, {Weiner}, {West}, {White}, {Wood-Vasey}, {Yeche}, {Zehavi},
  {Zhao}, \& {Zheng}}]{2013AJ....145...10D}
{Dawson}, K.~S., {Schlegel}, D.~J., {Ahn}, C.~P., {et~al.} 2013, \aj, 145, 10

\bibitem[{{de Jong} {et~al.}(2019){de Jong}, {Agertz}, {Berbel}, {Aird},
  {Alexander}, {Amarsi}, {Anders}, {Andrae}, {Ansarinejad}, {Ansorge},
  {Antilogus}, {Anwand-Heerwart}, {Arentsen}, {Arnadottir}, {Asplund}, {Auger},
  {Azais}, {Baade}, {Baker}, {Baker}, {Balbinot}, {Baldry}, {Banerji},
  {Barden}, {Barklem}, {Barth{\'e}l{\'e}my-Mazot}, {Battistini}, {Bauer},
  {Bell}, {Bellido-Tirado}, {Bellstedt}, {Belokurov}, {Bensby}, {Bergemann},
  {Bestenlehner}, {Bielby}, {Bilicki}, {Blake}, {Bland-Hawthorn}, {Boeche},
  {Boland}, {Boller}, {Bongard}, {Bongiorno}, {Bonifacio}, {Boudon}, {Brooks},
  {Brown}, {Brown}, {Br{\"u}ggen}, {Brynnel}, {Brzeski}, {Buchert},
  {Buschkamp}, {Caffau}, {Caillier}, {Carrick}, {Casagrande}, {Case}, {Casey},
  {Cesarini}, {Cescutti}, {Chapuis}, {Chiappini}, {Childress}, {Christlieb},
  {Church}, {Cioni}, {Cluver}, {Colless}, {Collett}, {Comparat}, {Cooper},
  {Couch}, {Courbin}, {Croom}, {Croton}, {Daguis{\'e}}, {Dalton}, {Davies},
  {Davis}, {de Laverny}, {Deason}, {Dionies}, {Disseau}, {Doel}, {D{\"o}scher},
  {Driver}, {Dwelly}, {Eckert}, {Edge}, {Edvardsson}, {Youssoufi}, {Elhaddad},
  {Enke}, {Erfanianfar}, {Farrell}, {Fechner}, {Feiz}, {Feltzing}, {Ferreras},
  {Feuerstein}, {Feuillet}, {Finoguenov}, {Ford}, {Fotopoulou}, {Fouesneau},
  {Frenk}, {Frey}, {Gaessler}, {Geier}, {Gentile Fusillo}, {Gerhard},
  {Giannantonio}, {Giannone}, {Gibson}, {Gillingham},
  {Gonz{\'a}lez-Fern{\'a}ndez}, {Gonzalez-Solares}, {Gottloeber}, {Gould},
  {Grebel}, {Gueguen}, {Guiglion}, {Haehnelt}, {Hahn}, {Hansen}, {Hartman},
  {Hauptner}, {Hawkins}, {Haynes}, {Haynes}, {Heiter}, {Helmi}, {Aguayo},
  {Hewett}, {Hinton}, {Hobbs}, {Hoenig}, {Hofman}, {Hook}, {Hopgood},
  {Hopkins}, {Hourihane}, {Howes}, {Howlett}, {Huet}, {Irwin}, {Iwert},
  {Jablonka}, {Jahn}, {Jahnke}, {Jarno}, {Jin}, {Jofre}, {Johl}, {Jones},
  {J{\"o}nsson}, {Jordan}, {Karovicova}, {Khalatyan}, {Kelz}, {Kennicutt},
  {King}, {Kitaura}, {Klar}, {Klauser}, {Kneib}, {Koch}, {Koposov},
  {Kordopatis}, {Korn}, {Kosmalski}, {Kotak}, {Kovalev}, {Kreckel}, {Kripak},
  {Krumpe}, {Kuijken}, {Kunder}, {Kushniruk}, {Lam}, {Lamer}, {Laurent},
  {Lawrence}, {Lehmitz}, {Lemasle}, {Lewis}, {Li}, {Lidman}, {Lind}, {Liske},
  {Lizon}, {Loveday}, {Ludwig}, {McDermid}, {Maguire}, {Mainieri}, {Mali},
  {Mandel}, {Mandel}, {Mannering}, {Martell}, {Martinez Delgado}, {Matijevic},
  {McGregor}, {McMahon}, {McMillan}, {Mena}, {Merloni}, {Meyer}, {Michel},
  {Micheva}, {Migniau}, {Minchev}, {Monari}, {Muller}, {Murphy},
  {Muthukrishna}, {Nandra}, {Navarro}, {Ness}, {Nichani}, {Nichol}, {Nicklas},
  {Niederhofer}, {Norberg}, {Obreschkow}, {Oliver}, {Owers}, {Pai},
  {Pankratow}, {Parkinson}, {Paschke}, {Paterson}, {Pecontal}, {Parry},
  {Phillips}, {Pillepich}, {Pinard}, {Pirard}, {Piskunov}, {Plank},
  {Pl{\"u}schke}, {Pons}, {Popesso}, {Power}, {Pragt}, {Pramskiy}, {Pryer},
  {Quattri}, {Queiroz}, {Quirrenbach}, {Rahurkar}, {Raichoor}, {Ramstedt},
  {Rau}, {Recio-Blanco}, {Reiss}, {Renaud}, {Revaz}, {Rhode}, {Richard},
  {Richter}, {Rix}, {Robotham}, {Roelfsema}, {Romaniello}, {Rosario},
  {Rothmaier}, {Roukema}, {Ruchti}, {Rupprecht}, {Rybizki}, {Ryde}, {Saar},
  {Sadler}, {Sahl{\'e}n}, {Salvato}, {Sassolas}, {Saunders}, {Saviauk},
  {Sbordone}, {Schmidt}, {Schnurr}, {Scholz}, {Schwope}, {Seifert}, {Shanks},
  {Sheinis}, {Sivov}, {Sk{\'u}lad{\'o}ttir}, {Smartt}, {Smedley}, {Smith},
  {Smith}, {Sorce}, {Spitler}, {Starkenburg}, {Steinmetz}, {Stilz}, {Storm},
  {Sullivan}, {Sutherland}, {Swann}, {Tamone}, {Taylor}, {Teillon}, {Tempel},
  {ter Horst}, {Thi}, {Tolstoy}, {Trager}, {Traven}, {Tremblay}, {Tresse},
  {Valentini}, {van de Weygaert}, {van den Ancker}, {Veljanoski}, {Venkatesan},
  {Wagner}, {Wagner}, {Walcher}, {Waller}, {Walton}, {Wang}, {Winkler},
  {Wisotzki}, {Worley}, {Worseck}, {Xiang}, {Xu}, {Yong}, {Zhao}, {Zheng},
  {Zscheyge}, \& {Zucker}}]{2019Msngr.175....3D}
{de Jong}, R.~S., {Agertz}, O., {Berbel}, A.~A., {et~al.} 2019, The Messenger,
  175, 3

\bibitem[{{Fern{\'a}ndez-Soto} {et~al.}(1999){Fern{\'a}ndez-Soto}, {Lanzetta},
  \& {Yahil}}]{1999ApJ...513...34F}
{Fern{\'a}ndez-Soto}, A., {Lanzetta}, K.~M., \& {Yahil}, A. 1999, \apj, 513, 34

\bibitem[{{Finkbeiner} {et~al.}(2016){Finkbeiner}, {Schlafly}, {Schlegel},
  {Padmanabhan}, {Juri{\'c}}, {Burgett}, {Chambers}, {Denneau}, {Draper},
  {Flewelling}, {Hodapp}, {Kaiser}, {Magnier}, {Metcalfe}, {Morgan}, {Price},
  {Stubbs}, \& {Tonry}}]{2016ApJ...822...66F}
{Finkbeiner}, D.~P., {Schlafly}, E.~F., {Schlegel}, D.~J., {et~al.} 2016, \apj,
  822, 66

\bibitem[{{Foreman-Mackey} {et~al.}(2013){Foreman-Mackey}, {Hogg}, {Lang}, \&
  {Goodman}}]{2013PASP..125..306F}
{Foreman-Mackey}, D., {Hogg}, D.~W., {Lang}, D., \& {Goodman}, J. 2013, \pasp,
  125, 306

\bibitem[{{Gavazzi} {et~al.}(2014){Gavazzi}, {Marshall}, {Treu}, \&
  {Sonnenfeld}}]{Gav++14}
{Gavazzi}, R., {Marshall}, P.~J., {Treu}, T., \& {Sonnenfeld}, A. 2014, \apj,
  785, 144

\bibitem[{{Gilman} {et~al.}(2020){Gilman}, {Birrer}, {Nierenberg}, {Treu},
  {Du}, \& {Benson}}]{Gil++20}
{Gilman}, D., {Birrer}, S., {Nierenberg}, A., {et~al.} 2020, \mnras, 491, 6077

\bibitem[{{Hezaveh} {et~al.}(2016){Hezaveh}, {Dalal}, {Marrone}, {Mao},
  {Morningstar}, {Wen}, {Blandford}, {Carlstrom}, {Fassnacht}, {Holder},
  {Kemball}, {Marshall}, {Murray}, {Perreault Levasseur}, {Vieira}, \&
  {Wechsler}}]{Hez++16}
{Hezaveh}, Y.~D., {Dalal}, N., {Marrone}, D.~P., {et~al.} 2016, \apj, 823, 37

\bibitem[{{Hildebrandt} {et~al.}(2021){Hildebrandt}, {van den Busch}, {Wright},
  {Blake}, {Joachimi}, {Kuijken}, {Tr{\"o}ster}, {Asgari}, {Bilicki}, {de
  Jong}, {Dvornik}, {Erben}, {Getman}, {Giblin}, {Heymans}, {Kannawadi}, {Lin},
  \& {Shan}}]{Hil++21}
{Hildebrandt}, H., {van den Busch}, J.~L., {Wright}, A.~H., {et~al.} 2021,
  \aap, 647, A124

\bibitem[{{Ilbert} {et~al.}(2010){Ilbert}, {Salvato}, {Le Floc'h}, {Aussel},
  {Capak}, {McCracken}, {Mobasher}, {Kartaltepe}, {Scoville}, {Sanders},
  {Arnouts}, {Bundy}, {Cassata}, {Kneib}, {Koekemoer}, {Le F{\`e}vre}, {Lilly},
  {Surace}, {Taniguchi}, {Tasca}, {Thompson}, {Tresse}, {Zamojski}, {Zamorani},
  \& {Zucca}}]{2010ApJ...709..644I}
{Ilbert}, O., {Salvato}, M., {Le Floc'h}, E., {et~al.} 2010, \apj, 709, 644

\bibitem[{{Jacobs} {et~al.}(2019){Jacobs}, {Collett}, {Glazebrook},
  {Buckley-Geer}, {Diehl}, {Lin}, {McCarthy}, {Qin}, {Odden}, {Caso Escudero},
  {Dial}, {Yung}, {Gaitsch}, {Pellico}, {Lindgren}, {Abbott}, {Annis}, {Avila},
  {Brooks}, {Burke}, {Carnero Rosell}, {Carrasco Kind}, {Carretero}, {da
  Costa}, {De Vicente}, {Fosalba}, {Frieman}, {Garc{\'\i}a-Bellido},
  {Gaztanaga}, {Goldstein}, {Gruen}, {Gruendl}, {Gschwend}, {Hollowood},
  {Honscheid}, {Hoyle}, {James}, {Krause}, {Kuropatkin}, {Lahav}, {Lima},
  {Maia}, {Marshall}, {Miquel}, {Plazas}, {Roodman}, {Sanchez}, {Scarpine},
  {Serrano}, {Sevilla-Noarbe}, {Smith}, {Sobreira}, {Suchyta}, {Swanson},
  {Tarle}, {Vikram}, {Walker}, {Zhang}, \& {DES Collaboration}}]{Jac++19}
{Jacobs}, C., {Collett}, T., {Glazebrook}, K., {et~al.} 2019, \apjs, 243, 17

\bibitem[{{Kawanomoto} {et~al.}(2018){Kawanomoto}, {Uraguchi}, {Komiyama},
  {Miyazaki}, {Furusawa}, {Finet}, {Hattori}, {Wang}, {Yasuda}, \&
  {Suzuki}}]{2018PASJ...70...66K}
{Kawanomoto}, S., {Uraguchi}, F., {Komiyama}, Y., {et~al.} 2018, \pasj, 70, 66

\bibitem[{{Kelvin} {et~al.}(2012){Kelvin}, {Driver}, {Robotham}, {Hill},
  {Alpaslan}, {Baldry}, {Bamford}, {Bland-Hawthorn}, {Brough}, {Graham},
  {H{\"a}ussler}, {Hopkins}, {Liske}, {Loveday}, {Norberg}, {Phillipps},
  {Popescu}, {Prescott}, {Taylor}, \& {Tuffs}}]{2012MNRAS.421.1007K}
{Kelvin}, L.~S., {Driver}, S.~P., {Robotham}, A. S.~G., {et~al.} 2012, \mnras,
  421, 1007

\bibitem[{{Kinney} {et~al.}(1996){Kinney}, {Calzetti}, {Bohlin}, {McQuade},
  {Storchi-Bergmann}, \& {Schmitt}}]{1996ApJ...467...38K}
{Kinney}, A.~L., {Calzetti}, D., {Bohlin}, R.~C., {et~al.} 1996, \apj, 467, 38

\bibitem[{{Kormann} {et~al.}(1994){Kormann}, {Schneider}, \&
  {Bartelmann}}]{1994A&A...284..285K}
{Kormann}, R., {Schneider}, P., \& {Bartelmann}, M. 1994, \aap, 284, 285

\bibitem[{{Kormendy} {et~al.}(2009){Kormendy}, {Fisher}, {Cornell}, \&
  {Bender}}]{2009ApJS..182..216K}
{Kormendy}, J., {Fisher}, D.~B., {Cornell}, M.~E., \& {Bender}, R. 2009, \apjs,
  182, 216

\bibitem[{{Kuijken} {et~al.}(2015){Kuijken}, {Heymans}, {Hildebrandt},
  {Nakajima}, {Erben}, {de Jong}, {Viola}, {Choi}, {Hoekstra}, {Miller}, {van
  Uitert}, {Amon}, {Blake}, {Brouwer}, {Buddendiek}, {Conti}, {Eriksen},
  {Grado}, {Harnois-D{\'e}raps}, {Helmich}, {Herbonnet}, {Irisarri},
  {Kitching}, {Klaes}, {La Barbera}, {Napolitano}, {Radovich}, {Schneider},
  {Sif{\'o}n}, {Sikkema}, {Simon}, {Tudorica}, {Valentijn}, {Verdoes Kleijn},
  \& {van Waerbeke}}]{Kui++15}
{Kuijken}, K., {Heymans}, C., {Hildebrandt}, H., {et~al.} 2015, \mnras, 454,
  3500

\bibitem[{{Le F{\`e}vre} {et~al.}(2013){Le F{\`e}vre}, {Cassata}, {Cucciati},
  {Garilli}, {Ilbert}, {Le Brun}, {Maccagni}, {Moreau}, {Scodeggio}, {Tresse},
  {Zamorani}, {Adami}, {Arnouts}, {Bardelli}, {Bolzonella}, {Bondi},
  {Bongiorno}, {Bottini}, {Cappi}, {Charlot}, {Ciliegi}, {Contini}, {de la
  Torre}, {Foucaud}, {Franzetti}, {Gavignaud}, {Guzzo}, {Iovino}, {Lemaux},
  {L{\'o}pez-Sanjuan}, {McCracken}, {Marano}, {Marinoni}, {Mazure}, {Mellier},
  {Merighi}, {Merluzzi}, {Paltani}, {Pell{\`o}}, {Pollo}, {Pozzetti},
  {Scaramella}, {Tasca}, {Vergani}, {Vettolani}, {Zanichelli}, \&
  {Zucca}}]{2013A&A...559A..14L}
{Le F{\`e}vre}, O., {Cassata}, P., {Cucciati}, O., {et~al.} 2013, \aap, 559,
  A14

\bibitem[{{Lilly} {et~al.}(1995){Lilly}, {Le Fevre}, {Crampton}, {Hammer}, \&
  {Tresse}}]{1995ApJ...455...50L}
{Lilly}, S.~J., {Le Fevre}, O., {Crampton}, D., {Hammer}, F., \& {Tresse}, L.
  1995, \apj, 455, 50

\bibitem[{{Madau}(1995)}]{1995ApJ...441...18M}
{Madau}, P. 1995, \apj, 441, 18

\bibitem[{{Magnier} {et~al.}(2013){Magnier}, {Schlafly}, {Finkbeiner}, {Juric},
  {Tonry}, {Burgett}, {Chambers}, {Flewelling}, {Kaiser}, {Kudritzki},
  {Morgan}, {Price}, {Sweeney}, \& {Stubbs}}]{ps13pi}
{Magnier}, E.~A., {Schlafly}, E., {Finkbeiner}, D., {et~al.} 2013, \apjs, 205,
  20

\bibitem[{{Momcheva} {et~al.}(2016){Momcheva}, {Brammer}, {van Dokkum},
  {Skelton}, {Whitaker}, {Nelson}, {Fumagalli}, {Maseda}, {Leja}, {Franx},
  {Rix}, {Bezanson}, {Da Cunha}, {Dickey}, {F{\"o}rster Schreiber},
  {Illingworth}, {Kriek}, {Labb{\'e}}, {Ulf Lange}, {Lundgren}, {Magee},
  {Marchesini}, {Oesch}, {Pacifici}, {Patel}, {Price}, {Tal}, {Wake}, {van der
  Wel}, \& {Wuyts}}]{2016ApJS..225...27M}
{Momcheva}, I.~G., {Brammer}, G.~B., {van Dokkum}, P.~G., {et~al.} 2016, \apjs,
  225, 27

\bibitem[{{Myles} {et~al.}(2021){Myles}, {Alarcon}, {Amon}, {S{\'a}nchez},
  {Everett}, {DeRose}, {McCullough}, {Gruen}, {Bernstein}, {Troxel},
  {Dodelson}, {Campos}, {MacCrann}, {Yin}, {Raveri}, {Amara}, {Becker}, {Choi},
  {Cordero}, {Eckert}, {Gatti}, {Giannini}, {Gschwend}, {Gruendl}, {Harrison},
  {Hartley}, {Huff}, {Kuropatkin}, {Lin}, {Masters}, {Miquel}, {Prat},
  {Roodman}, {Rykoff}, {Sevilla-Noarbe}, {Sheldon}, {Wechsler}, {Yanny},
  {Abbott}, {Aguena}, {Allam}, {Annis}, {Bacon}, {Bertin}, {Bhargava},
  {Bridle}, {Brooks}, {Burke}, {Carnero Rosell}, {Carrasco Kind}, {Carretero},
  {Castander}, {Conselice}, {Costanzi}, {Crocce}, {da Costa}, {Pereira},
  {Desai}, {Diehl}, {Eifler}, {Elvin-Poole}, {Evrard}, {Ferrero}, {Fert{\'e}},
  {Flaugher}, {Fosalba}, {Frieman}, {Garc{\'\i}a-Bellido}, {Gaztanaga},
  {Giannantonio}, {Hinton}, {Hollowood}, {Honscheid}, {Hoyle}, {Huterer},
  {James}, {Krause}, {Kuehn}, {Lahav}, {Lima}, {Maia}, {Marshall}, {Martini},
  {Melchior}, {Menanteau}, {Mohr}, {Morgan}, {Muir}, {Ogando}, {Palmese},
  {Paz-Chinch{\'o}n}, {Plazas}, {Rodriguez-Monroy}, {Samuroff}, {Sanchez},
  {Scarpine}, {Secco}, {Serrano}, {Smith}, {Soares-Santos}, {Suchyta},
  {Swanson}, {Tarle}, {Thomas}, {To}, {Varga}, {Weller}, \& {Wester}}]{Myl++21}
{Myles}, J., {Alarcon}, A., {Amon}, A., {et~al.} 2021, \mnras, 505, 4249

\bibitem[{{Newman} {et~al.}(2015){Newman}, {Ellis}, \& {Treu}}]{New++15}
{Newman}, A.~B., {Ellis}, R.~S., \& {Treu}, T. 2015, \apj, 814, 26

\bibitem[{{Oguri}(2006)}]{2006MNRAS.367.1241O}
{Oguri}, M. 2006, \mnras, 367, 1241

\bibitem[{{Oldham} \& {Auger}(2018)}]{O+A18}
{Oldham}, L.~J. \& {Auger}, M.~W. 2018, \mnras, 476, 133

\bibitem[{{Petrillo} {et~al.}(2019){Petrillo}, {Tortora}, {Vernardos},
  {Koopmans}, {Verdoes Kleijn}, {Bilicki}, {Napolitano}, {Chatterjee},
  {Covone}, {Dvornik}, {Erben}, {Getman}, {Giblin}, {Heymans}, {de Jong},
  {Kuijken}, {Schneider}, {Shan}, {Spiniello}, \& {Wright}}]{Pet++19}
{Petrillo}, C.~E., {Tortora}, C., {Vernardos}, G., {et~al.} 2019, \mnras, 484,
  3879

\bibitem[{{Rojas} {et~al.}(2021){Rojas}, {Savary}, {Cl{\'e}ment}, {Maus},
  {Courbin}, {Lemon}, {Chan}, {Vernardos}, {Joseph}, {Ca{\~n}ameras}, \&
  {Galan}}]{Roj++21}
{Rojas}, K., {Savary}, E., {Cl{\'e}ment}, B., {et~al.} 2021, arXiv e-prints,
  arXiv:2109.00014

\bibitem[{{Ruff} {et~al.}(2011){Ruff}, {Gavazzi}, {Marshall}, {Treu}, {Auger},
  \& {Brault}}]{Ruf++11}
{Ruff}, A.~J., {Gavazzi}, R., {Marshall}, P.~J., {et~al.} 2011, \apj, 727, 96

\bibitem[{{Salvato} {et~al.}(2019){Salvato}, {Ilbert}, \& {Hoyle}}]{SIH19}
{Salvato}, M., {Ilbert}, O., \& {Hoyle}, B. 2019, Nature Astronomy, 3, 212

\bibitem[{{Sawicki} {et~al.}(1997){Sawicki}, {Lin}, \&
  {Yee}}]{1997AJ....113....1S}
{Sawicki}, M.~J., {Lin}, H., \& {Yee}, H.~K.~C. 1997, \aj, 113, 1

\bibitem[{{Schlafly} {et~al.}(2012){Schlafly}, {Finkbeiner}, {Juri{\'c}},
  {Magnier}, {Burgett}, {Chambers}, {Grav}, {Hodapp}, {Kaiser}, {Kudritzki},
  {Martin}, {Morgan}, {Price}, {Rix}, {Stubbs}, {Tonry}, \&
  {Wainscoat}}]{2012ApJ...756..158S}
{Schlafly}, E.~F., {Finkbeiner}, D.~P., {Juri{\'c}}, M., {et~al.} 2012, \apj,
  756, 158

\bibitem[{{Schlegel} {et~al.}(2019){Schlegel}, {Kollmeier}, \&
  {Ferraro}}]{2019BAAS...51g.229S}
{Schlegel}, D., {Kollmeier}, J.~A., \& {Ferraro}, S. 2019, in Bulletin of the
  American Astronomical Society, Vol.~51, 229

\bibitem[{{Schuldt} {et~al.}(2019){Schuldt}, {Chiriv{\`\i}}, {Suyu},
  {Y{\i}ld{\i}r{\i}m}, {Sonnenfeld}, {Halkola}, \& {Lewis}}]{Sch++19}
{Schuldt}, S., {Chiriv{\`\i}}, G., {Suyu}, S.~H., {et~al.} 2019, \aap, 631, A40

\bibitem[{{Sersic}(1968)}]{1968adga.book.....S}
{Sersic}, J.~L. 1968, {Atlas de Galaxias Australes}

\bibitem[{{Shajib} {et~al.}(2021){Shajib}, {Treu}, {Birrer}, \&
  {Sonnenfeld}}]{Sha++21}
{Shajib}, A.~J., {Treu}, T., {Birrer}, S., \& {Sonnenfeld}, A. 2021, \mnras,
  503, 2380

\bibitem[{{Skelton} {et~al.}(2014){Skelton}, {Whitaker}, {Momcheva}, {Brammer},
  {van Dokkum}, {Labb{\'e}}, {Franx}, {van der Wel}, {Bezanson}, {Da Cunha},
  {Fumagalli}, {F{\"o}rster Schreiber}, {Kriek}, {Leja}, {Lundgren}, {Magee},
  {Marchesini}, {Maseda}, {Nelson}, {Oesch}, {Pacifici}, {Patel}, {Price},
  {Rix}, {Tal}, {Wake}, \& {Wuyts}}]{2014ApJS..214...24S}
{Skelton}, R.~E., {Whitaker}, K.~E., {Momcheva}, I.~G., {et~al.} 2014, \apjs,
  214, 24

\bibitem[{{Smith} {et~al.}(2015){Smith}, {Lucey}, \& {Conroy}}]{SLC15}
{Smith}, R.~J., {Lucey}, J.~R., \& {Conroy}, C. 2015, \mnras, 449, 3441

\bibitem[{{Sonnenfeld}(2022{\natexlab{a}})}]{Son22a}
{Sonnenfeld}, A. 2022{\natexlab{a}}, \aap, 659, A132

\bibitem[{{Sonnenfeld}(2022{\natexlab{b}})}]{Son22b}
{Sonnenfeld}, A. 2022{\natexlab{b}}, \aap, 659, A133

\bibitem[{{Sonnenfeld} {et~al.}(2018){Sonnenfeld}, {Chan}, {Shu}, {More},
  {Oguri}, {Suyu}, {Wong}, {Lee}, {Coupon}, {Yonehara}, {Bolton}, {Jaelani},
  {Tanaka}, {Miyazaki}, \& {Komiyama}}]{2018PASJ...70S..29S}
{Sonnenfeld}, A., {Chan}, J. H.~H., {Shu}, Y., {et~al.} 2018, \pasj, 70, S29

\bibitem[{{Sonnenfeld} {et~al.}(2019{\natexlab{a}}){Sonnenfeld}, {Jaelani},
  {Chan}, {More}, {Suyu}, {Wong}, {Oguri}, \& {Lee}}]{Son++19}
{Sonnenfeld}, A., {Jaelani}, A.~T., {Chan}, J., {et~al.} 2019{\natexlab{a}},
  \aap, 630, A71

\bibitem[{{Sonnenfeld} {et~al.}(2012){Sonnenfeld}, {Treu}, {Gavazzi},
  {Marshall}, {Auger}, {Suyu}, {Koopmans}, \& {Bolton}}]{Son++12}
{Sonnenfeld}, A., {Treu}, T., {Gavazzi}, R., {et~al.} 2012, \apj, 752, 163

\bibitem[{{Sonnenfeld} {et~al.}(2013){Sonnenfeld}, {Treu}, {Gavazzi}, {Suyu},
  {Marshall}, {Auger}, \& {Nipoti}}]{Son++13b}
{Sonnenfeld}, A., {Treu}, T., {Gavazzi}, R., {et~al.} 2013, \apj, 777, 98

\bibitem[{{Sonnenfeld} {et~al.}(2015){Sonnenfeld}, {Treu}, {Marshall}, {Suyu},
  {Gavazzi}, {Auger}, \& {Nipoti}}]{Son++15}
{Sonnenfeld}, A., {Treu}, T., {Marshall}, P.~J., {et~al.} 2015, \apj, 800, 94

\bibitem[{{Sonnenfeld} {et~al.}(2020){Sonnenfeld}, {Verma}, {More}, {Baeten},
  {Macmillan}, {Wong}, {Chan}, {Jaelani}, {Lee}, {Oguri}, {Rusu}, {Veldthuis},
  {Trouille}, {Marshall}, {Hutchings}, {Allen}, {O'Donnell}, {Cornen}, {Davis},
  {McMaster}, {Lintott}, \& {Miller}}]{2020A&A...642A.148S}
{Sonnenfeld}, A., {Verma}, A., {More}, A., {et~al.} 2020, \aap, 642, A148

\bibitem[{{Sonnenfeld} {et~al.}(2019{\natexlab{b}}){Sonnenfeld}, {Wang}, \&
  {Bahcall}}]{2019A&A...622A..30S}
{Sonnenfeld}, A., {Wang}, W., \& {Bahcall}, N. 2019{\natexlab{b}}, \aap, 622,
  A30

\bibitem[{{Storfer} {et~al.}(2022){Storfer}, {Huang}, {Gu}, {Sheu}, {Banka},
  {Dey}, {Jain}, {Kwon}, {Lang}, {Lee}, {Meisner}, {Moustakas}, {Myers},
  {Tabares-Tarquinio}, {Schlafly}, \& {Schlegel}}]{2022arXiv220602764S}
{Storfer}, C., {Huang}, X., {Gu}, A., {et~al.} 2022, arXiv e-prints,
  arXiv:2206.02764

\bibitem[{{Tanaka} {et~al.}(2018){Tanaka}, {Coupon}, {Hsieh}, {Mineo},
  {Nishizawa}, {Speagle}, {Furusawa}, {Miyazaki}, \& {Murayama}}]{Tan++18}
{Tanaka}, M., {Coupon}, J., {Hsieh}, B.-C., {et~al.} 2018, \pasj, 70, S9

\bibitem[{{The MSE Science Team} {et~al.}(2019){The MSE Science Team},
  {Babusiaux}, {Bergemann}, {Burgasser}, {Ellison}, {Haggard}, {Huber},
  {Kaplinghat}, {Li}, {Marshall}, {Martell}, {McConnachie}, {Percival},
  {Robotham}, {Shen}, {Thirupathi}, {Tran}, {Yeche}, {Yong}, {Adibekyan},
  {Silva Aguirre}, {Angelou}, {Asplund}, {Balogh}, {Banerjee}, {Bannister},
  {Barr{\'\i}a}, {Battaglia}, {Bayo}, {Bechtol}, {Beck}, {Beers}, {Bellinger},
  {Berg}, {Bestenlehner}, {Bilicki}, {Bitsch}, {Bland-Hawthorn}, {Bolton},
  {Boselli}, {Bovy}, {Bragaglia}, {Buzasi}, {Caffau}, {Cami}, {Carleton},
  {Casagrande}, {Cassisi}, {Catelan}, {Chang}, {Cortese}, {Damjanov}, {Davies},
  {de Grijs}, {de Rosa}, {Deason}, {di Matteo}, {Drlica-Wagner}, {Erkal},
  {Escorza}, {Ferrarese}, {Fleming}, {Font-Ribera}, {Freeman}, {G{\"a}nsicke},
  {Gabdeev}, {Gallagher}, {Gandolfi}, {Garc{\'\i}a}, {Gaulme}, {Geha},
  {Gennaro}, {Gieles}, {Gilbert}, {Gordon}, {Goswami}, {Greco}, {Grillmair},
  {Guiglion}, {H{\'e}nault-Brunet}, {Hall}, {Handler}, {Hansen}, {Hathi},
  {Hatzidimitriou}, {Haywood}, {Hern{\'a}ndez Santisteban}, {Hillenbrand},
  {Hopkins}, {Howlett}, {Hudson}, {Ibata}, {Ili{\'c}}, {Jablonka}, {Ji},
  {Jiang}, {Juneau}, {Karakas}, {Karinkuzhi}, {Kim}, {Kong}, {Konstantopoulos},
  {Krogager}, {Lagos}, {Lallement}, {Laporte}, {Lebreton}, {Lee}, {Lewis},
  {Lianou}, {Liu}, {Lodieu}, {Loveday}, {M{\'e}sz{\'a}ros}, {Makler}, {Mao},
  {Marchesini}, {Martin}, {Mateo}, {Melis}, {Merle}, {Miglio}, {Gohar
  Mohammad}, {Molaverdikhani}, {Monier}, {Morel}, {Mosser}, {Nataf}, {Necib},
  {Neilson}, {Newman}, {Nierenberg}, {Nord}, {Noterdaeme}, {O'Dea}, {Oshagh},
  {Pace}, {Palanque-Delabrouille}, {Pandey}, {Parker}, {Pawlowski}, {Peter},
  {Petitjean}, {Petric}, {Placco}, {Popovi{\'c}}, {Price-Whelan}, {Prsa},
  {Ravindranath}, {Rich}, {Ruan}, {Rybizki}, {Sakari}, {Sanderson}, {Schiavon},
  {Schimd}, {Serenelli}, {Siebert}, {Siudek}, {Smiljanic}, {Smith}, {Sobeck},
  {Starkenburg}, {Stello}, {Szab{\'o}}, {Szabo}, {Taylor}, {Thanjavur},
  {Thomas}, {Tollerud}, {Toonen}, {Tremblay}, {Tresse}, {Tsantaki},
  {Valentini}, {Van Eck}, {Variu}, {Venn}, {Villaver}, {Walker}, {Wang},
  {Wang}, {Wilson}, {Wright}, {Xu}, {Yildiz}, {Zhang}, {Zwintz}, {Anguiano},
  {Bedell}, {Chaplin}, {Collet}, {Cuillandre}, {Duc}, {Flagey}, {Hermes},
  {Hill}, {Kamath}, {Laychak}, {Ma{\l}ek}, {Marley}, {Sheinis}, {Simons},
  {Sousa}, {Szeto}, {Ting}, {Vegetti}, {Wells}, {Babas}, {Bauman}, {Bosselli},
  {C{\^o}t{\'e}}, {Colless}, {Comparat}, {Courtois}, {Crampton}, {Croom},
  {Davies}, {de Grijs}, {Denny}, {Devost}, {di Matteo}, {Driver},
  {Fernandez-Lorenzo}, {Guhathakurta}, {Han}, {Higgs}, {Hill}, {Ho}, {Hopkins},
  {Hudson}, {Ibata}, {Isani}, {Jarvis}, {Johnson}, {Jullo}, {Kaiser}, {Kneib},
  {Koda}, {Koshy}, {Mignot}, {Murowinski}, {Newman}, {Nusser}, {Pancoast},
  {Peng}, {Peroux}, {Pichon}, {Poggianti}, {Richard}, {Salmon}, {Seibert},
  {Shastri}, {Smith}, {Sutaria}, {Tao}, {Taylor}, {Tully}, {van Waerbeke},
  {Vermeulen}, {Walker}, {Willis}, {Willot}, \&
  {Withington}}]{2019arXiv190404907T}
{The MSE Science Team}, {Babusiaux}, C., {Bergemann}, M., {et~al.} 2019, arXiv
  e-prints, arXiv:1904.04907

\bibitem[{{Tran} {et~al.}(2022){Tran}, {Harshan}, {Glazebrook}, {Keerthi
  Vasan}, {Jones}, {Jacobs}, {Kacprzak}, {Barone}, {Collett}, {Gupta},
  {Henderson}, {Kewley}, {Lopez}, {Nanayakkara}, {Sanders}, \&
  {Sweet}}]{2022AJ....164..148T}
{Tran}, K.-V.~H., {Harshan}, A., {Glazebrook}, K., {et~al.} 2022, \aj, 164, 148

\bibitem[{{Treu} {et~al.}(2010){Treu}, {Auger}, {Koopmans}, {Gavazzi},
  {Marshall}, \& {Bolton}}]{Tre++10}
{Treu}, T., {Auger}, M.~W., {Koopmans}, L. V.~E., {et~al.} 2010, \apj, 709,
  1195

\bibitem[{{Treu} \& {Koopmans}(2004)}]{T+K04}
{Treu}, T. \& {Koopmans}, L. V.~E. 2004, \apj, 611, 739

\bibitem[{{Vegetti} {et~al.}(2010){Vegetti}, {Koopmans}, {Bolton}, {Treu}, \&
  {Gavazzi}}]{Veg++10}
{Vegetti}, S., {Koopmans}, L.~V.~E., {Bolton}, A., {Treu}, T., \& {Gavazzi}, R.
  2010, \mnras, 408, 1969

\bibitem[{{Williams} {et~al.}(1996){Williams}, {Blacker}, {Dickinson}, {Dixon},
  {Ferguson}, {Fruchter}, {Giavalisco}, {Gilliland}, {Heyer}, {Katsanis},
  {Levay}, {Lucas}, {McElroy}, {Petro}, {Postman}, {Adorf}, \&
  {Hook}}]{1996AJ....112.1335W}
{Williams}, R.~E., {Blacker}, B., {Dickinson}, M., {et~al.} 1996, \aj, 112,
  1335

\bibitem[{{Wong} {et~al.}(2018){Wong}, {Sonnenfeld}, {Chan}, {Rusu}, {Tanaka},
  {Jaelani}, {Lee}, {More}, {Oguri}, {Suyu}, \&
  {Komiyama}}]{2018ApJ...867..107W}
{Wong}, K.~C., {Sonnenfeld}, A., {Chan}, J. H.~H., {et~al.} 2018, \apj, 867,
  107

\end{thebibliography}
\bibliographystyle{aa}

\begin{appendix}

\section{Lens residuals}
\label{sect:lensres}

Equation \ref{equation: omega vs delta c} can be used to predict the $g-i$ colour measurement error, $\delta C_{gi}$, for each best-fit model to the simulated systems. Here, we compare these predicted $\delta C_{gi}$ against the true $\delta C_{gi}$ of the best-fit models (measured in Section \ref{colour measurement}). We start by re-writing  Equation \ref{equation: omega vs delta c}. Defining the residual ratio parameter $\omega$ as 
\begin{equation}
    \omega = \log_{10}\frac{1+\psi_g}{1+\psi_i}\;,
    \label{A - 5}
\end{equation}
Equation \ref{equation: omega vs delta c} can be re-written as 
\begin{equation}
    \delta C_{gi} = -2.5 \omega\;.
    \label{A - equation: omega vs delta c}
\end{equation}
From Equation \ref{A - equation: omega vs delta c}, we  expect the colour measurement error to be inversely proportional to $\omega$. We want to compare this prediction with the results from our fit to the simulated lenses. 

Measuring $\omega$ on the best-fit models of Section \ref{colour measurement} requires specifying the definitions of the source image position and the lens residuals at the source image position. As in Section \ref{simulations}, we define the source footprint as the $g$-band source-only pixels with values higher than $3\sigma_{\text{sky}}$. With this definition, the source footprint can consist of multiple disconnected patches. We define the total lens residuals at the source image position in each band ($\Sigma_{\text{res,band}}$) as the sum of all the pixels of the lens residual image that lie within the source footprint (the lens residual image is produced by subtracting the best-fit lens light model from the lens original image). Similarly, we define the total source light in each band ($\Sigma_{\text{source,band}}$), as the sum of all the pixels of the source-only image that lie within the source footprint. Using these definitions, $\psi_{\textnormal{band}}$ in Equation \ref{psi text} can be re-written as 
\begin{equation}
    \psi_{\text{band}} \approx \frac{\Sigma_{\text{res,band}}}{\Sigma_{\text{source,band}}}\;,
    \label{A - 6}
\end{equation}
which allows   $\omega$ to be measured on the best-fit models. 

\begin{figure}
    \centering
    \includegraphics[width=9cm]{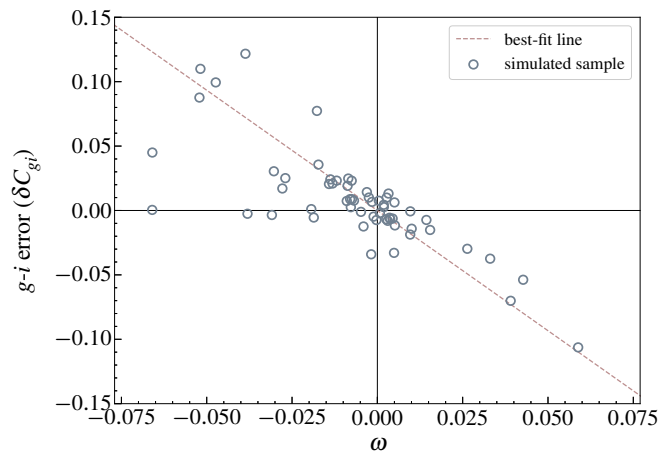}
    \caption{Source colour measurement error correlates with the lens light residuals. This figure shows the error of the measured source $g-i$ colours ($\delta C_{gi}$), plotted against the residual ratio parameter ($\omega$, defined in Equations \ref{A - 5} and \ref{A - 6}). The dashed line shows the best-fit line passing through the origin with a $-1.867 \pm 0.028$ slope. The thin grey lines show the $\omega=0$ and the $\delta C_{gi}=0$ lines.}
    \label{residuals}
\end{figure}

Figure \ref{residuals} shows the correlation between the $\delta C_{gi}$ and $\omega$.
We fitted a linear model with zero intercept to this distribution, finding a best-fit slope of $-1.867 \pm 0.028$ (dashed line). This correlation simply indicates that $|\delta C_{gi}|$ grows with the ratio of the lens residual flux to the source flux at the source image position, and is positive (negative) if the $g-i$ colour of the residuals is bigger (smaller) than the true source $g-i$ colour. The slope inferred from the best-fit models is shallower than the expected slope from Equation \ref{A - equation: omega vs delta c} (-2.5). This discrepancy is rooted in our initial assumption that all the lens residuals at the source image position get overfitted by the best-fit source light model (see Section \ref{section: the accuracy of the lens light model}). The weaker correlation probed here indicates that this is not necessarily the case, and that the -2.5 slope should be treated as the most conservative limit.

\section{Photometric redshift estimation of simulated isolated galaxies}
\label{app: isolated galaxies}

\begin{figure}
    \centering
    \includegraphics[width=9cm]{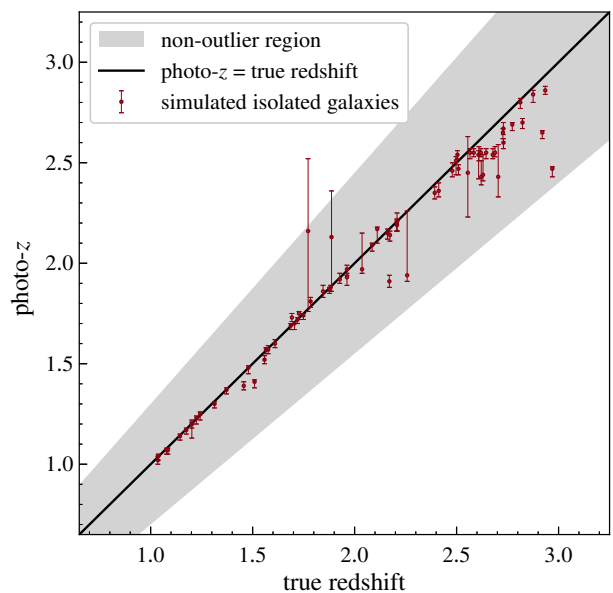} 
    \caption{Estimated photo-$z$s vs the true redshifts for our simulated isolated galaxies. The black line shows where photo-$z$s equal true redshifts. The error bars correspond to the 68th percentile of the redshift PDFs. The grey shaded region shows where $\delta z/(1+z_{\mathrm{true}}) < 0.15$. There are no photo-$z$ outliers, and the $1\sigma$ scatter of redshift errors ($\delta z/1+z_{\mathrm{true}}$) is 0.018.}
    \label{photoz: isolated}
\end{figure}

In Section \ref{results: simulated sample} we evaluate the performance of our algorithm in estimating the photo-$z$s of simulated strongly lensed galaxies. 
Here we use a simulated sample of isolated galaxies to evaluate the impact of lensing on the accuracy of photo-$z$ estimation.
For each strongly lensed system in our simulated sample (Section \ref{simulations}), we use the unlensed image of the source galaxy in each $grizy$ band as the simulated isolated galaxy. 
In each band, we convolve this image with the PSF of the lens galaxy that was originally assigned to this source in our simulated sample of strongly lensed systems (Section \ref{simulations}).
We do this to limit the variations between the sample of simulated isolated galaxies and the sample of simulated strongly lensed systems to lensing. 
In each band we centre the PSF-convolved image of the simulated isolated galaxy on the central pixel of a $101 \times 101$ pixel cutout.

We measure the colours by fitting a single S\'{e}rsic profile to each isolated galaxy, using a similar method to that   used in Section \ref{modelling the strongly lensed systems} to fit the surface brightness of lens galaxies. 
The best-fit $g-i$, $r-i$, $i-z$, and $i-y$ source colours have $\sim 0.004$ standard deviations and $\leq 0.0005$ bias around their true values. 
We estimate the photo-$z$s by running the BPZ algorithm with the best-fit colours and their uncertainties (0.004 for all the colours, see Section \ref{results: simulated sample}).
In order to isolate the variations between the experiment here and that of Section \ref{results: simulated sample} to lensing, as in  Section \ref{results: simulated sample}, we use a flat prior for the reference-band magnitude dependence of redshift.
Figure \ref{photoz: isolated} shows the estimated photo-$z$s for all the simulated isolated galaxies. The error bars show the 68th percentile of the redshift PDFs.
There are no outliers amongst the estimated photo-$z$s, and the $1\sigma$ scatter of the redshift errors ($\delta z/(1+z_{\mathrm{true}})$) is 0.018. In contrast, the photo-$z$ estimation of simulated strongly lensed galaxies had an $8.3\%$ outlier fraction and a 0.032 $1\sigma$ scatter of redshift errors (Section \ref{results: simulated sample}).

\end{appendix}

\end{document}